\documentclass[%
 amsmath,amssymb,
prf,
author-numerical,%
]{revtex4-1}
\usepackage{lineno,hyperref}
\usepackage{rotating, graphicx}
\usepackage{caption,subcaption}
\usepackage{color}

\graphicspath{{./graph/}}
\usepackage{color}

\begin{document}

\title{Effect of surfactant on elongated bubbles in capillary tubes at high Reynolds number}
\author{A.~Batchvarov$^1$}
\author{L.~Kahouadji$^1$}
\author{M.~Magnini$^2$}
\author{C.~R.~Constante-Amores$^1$}
\author{R.~V.~Craster$^3$}
\author{S.~Shin$^4$}
\author{J.~Chergui$^5$}
\author{D.~Juric$^5$}
\author{O.~K.~Matar$^1$}

\affiliation{$^1$Department of Chemical Engineering, Imperial College London, South Kensington Campus, London SW7 2AZ, United Kingdom \\}
\affiliation{$^2$Department of Mechanical, Materials and Manufacturing Engineering, University of Nottingham, Nottingham NG7 2RD, United Kingdom \\}
\affiliation{$^3$Department of Mathematics, Imperial College London, South Kensington Campus, London SW7 2AZ, United Kingdom \\}
\affiliation{$^4$Department of Mechanical and System Design Engineering, Hongik University, Seoul 121-791, Republic of Korea \\}
\affiliation{$^5$Laboratoire d'Informatique pour la M\'ecanique et les Sciences de l'Ing\'enieur (LIMSI), Centre National de la Recherche Scientifique (CNRS), Universit\'e Paris Saclay, B\^at. 507, Rue du Belv\'ed\`ere, Campus Universitaire, 91405 Orsay, France}

\date{\today}

\begin{abstract}
The effect of surfactants on the tail and film dynamics of elongated gas bubbles propagating through circular capillary tubes is investigated by means of an extensive three-dimensional numerical study using a hybrid front-tracking/level-set method. The focus is on the visco-inertial regime, which occurs when the Reynolds number of the flow is much larger than unity. Under these conditions, `clean' bubbles exhibit interface undulations in the proximity of the tail \cite{Magnini_prf_2017}, with an amplitude that increases with the Reynolds number. We perform a 
systematic analysis of the impact of a wide range of surfactant properties, including elasticity, bulk surfactant concentration, solubility, and diffusivity, on the bubble and flow dynamics in the presence of inertial effects. The results show that the introduction of surfactants is effective in suppressing the tail undulations as they tend to accumulate near the bubble tail. Here, large Marangoni stresses are generated, which lead to a local `rigidification' of the bubble. This effect becomes more pronounced for larger surfactant elasticities and adsorption depths. At reduced surfactant solubility, a thicker rigid film region forms at the bubble rear, where a Couette film flow is established, while undulations still appear at the trailing edge of the downstream `clean' film region. In such conditions, the bubble length becomes an influential parameter, with short bubbles becoming completely rigid.
\end{abstract}

\maketitle

\section{Introduction}
The dynamics of elongated bubbles within liquid-filled capillary channels is receiving increasing scientific and industrial attention due to its widespread occurrence in many processes such as modern electronic cooling systems, enhanced oil recovery, and coating processes, to name a few. When an elongated gas bubble is transported by liquid in a capillary tube, the front and rear caps are separated by a cylindrical region where the thickness of the liquid film trapped against the channel wall is uniform. Initial experimental work by Taylor \cite{Taylor_jfm_1960} showed that the film thickness ratio $h_0/R$ (where $R$ and $h_0$ refer to the tube radius and the uniform liquid film thickness, respectively) attains an asymptotic value of $1/3$ when the bubble capillary number $Ca_b = \mu U_b/ \sigma$ approaches  $2$ (with $\mu$ being the viscosity of the liquid, $\sigma$ the surface tension, and $U_b$ the bubble velocity). Using lubrication theory, Bretherton \cite{Bretherton_jfm_1961} showed that $h_0/R \approx Ca_b^{2/3}$  in the limits of  $Ca_b\ll1$ and $Re_b\ll 1$,  where $Re_b= 2\rho U_b R/\mu$ refers to the Reynolds number (with $\rho$ being the density of the liquid). 

Since the seminal works of Taylor \cite{Taylor_jfm_1960} and Bretherton \cite{Bretherton_jfm_1961}, the motion of long gas bubbles in capillary tubes has become a classical fluid mechanics problem and has been studied extensively. The effect of inertia on the front meniscus and film thickness of the bubble was studied experimentally by Aussillous and  Qu\'er\'e \cite{Aussillous_pof_2000}, who provided an empirical correction to Bretherton's law to fit Taylor's film thickness data  \cite{Taylor_jfm_1960} at large capillary numbers, and proposed a scaling law for the film thickness in the presence of inertia. Han and Shikazono \cite{Han_ijhff_2009} performed direct film thickness measurements with optical techniques using low viscosity fluids and, in the limit of $Re_b< 2000$, found that their experimental data were best correlated as 
\begin{equation}
	\frac{h_0}{R}=\frac{1.34 Ca_b^{2/3}}{1 +3.13 Ca_b^{2/3}+0.504 Ca_b^{0.672} Re_b^{0.589}-0.352 We_b^{0.629}}. 
	\label{HS}
	\end{equation}
\noindent where the Weber number was defined as $We_b=2 \rho U_b^2 R/\sigma$.

A lubrication theory approach to quantify film thickness in the presence of inertia was used by de Ryck \cite{deRyck_pof_2002}. Numerical studies of the behaviour of the front meniscus in the presence of inertia have looked into the meniscus shape and film thickness \cite{Giavedoni_pof_1997}, vortical structures ahead of the bubble tip \cite{Heil_pof_2001}, and pressure drop across the bubble \cite{Kreutzer_aiche_2005}. Numerical studies into the rear meniscus of the bubble have reported that bubble undulations appear in the proximity of the tail when $Re_b\gg1$ and become more apparent as $Re_b$ increases \cite{Edvinsson_aiche_1996,Giavedoni_pof_1999}. More recently, Magnini {\it et al.} \cite{Magnini_prf_2017} have used both lubrication theory and direct numerical simulation approaches to  study systematically the effect of $Re_b$ on the tail dynamics. They observed that increasing inertia  decreases monotonically the wavelength of the tail undulations, which was also observed experimentally by Khodaparast {\it et al.} \cite{Khodaparast_mano_2015}. 

Often bubble dynamics is affected by deliberately-placed or accidentally-found surface active agents. Surfactants find it more energy favourable to migrate toward fluid interfaces, where they act to reduce surface tension. The presence of non-uniform interfacial species concentration can lead to surface tension gradients, which, in turn, give rise to Marangoni stresses that drive fluid away from regions of high surfactant concentration \cite{Matar_sf_2009}. The presence of surfactants for confined gas-liquid systems plays a significant role on the re-opening of pulmonary airways \cite{Grotberg_arbe_2001,Halpern_rpn_2008,Heil_rpn_2008,Grotberg_pof_2011}, where the lack of surfactants can lead to higher surface tension at the air-liquid interface, leading to blockage of the passage way \cite{Heil_rpn_2008}.

The significance of surfactant effects on confined gas-liquid systems has led to a number of theoretical works, built upon the simplifying assumption of negligible inertia ($Re\ll 1$). Ginley and Radke \cite{Ginley_acsss_1989} studied the effect of adsorption controlled soluble surfactant transport on the motion of gas bubbles in cylindrical tubes and reported that their presence results in an increased pressure drop across the bubble and a slightly decreased thin film thickness. Ratulowski and Chang \cite{Ratulowski_jfm_1990}  have shown that the presence of surfactant bulk concentration gradients can act to increase the liquid film thickness by a maximum factor of $4^{2/3}$ in comparison to the Bretherton theory \cite{Bretherton_jfm_1961}. This result was later confirmed by works from Park \cite{Park_pofA_1992} and Stebe and Barth\`es-Biesel \cite{Stebe_jfm_1995} and was largely attributed to the presence of higher surfactant concentration at the front of the bubble in comparison to the thin film region, where Marangoni stresses act to drive fluid from the front into the thin film region. Further theoretical work by Borhan and Mao \cite{Borhan_PofA_1992} focused on the effect of insoluble surfactants on the motion and deformation of gas bubbles and reported that the presence of Marangoni stresses acts to retard the bubble motion by opposing surface convection. Experimental works on the effect of surfactants on film thickness have been primarily focused on coating processes. Ou Ramdane and Qu\'er\'e \cite{OuRamdane_Langmuir_1997} focused on fiber-coating and observed a film-thickening factor ranging between $1$ and $4^{2/3}$ compared to the `clean' interface case, depending on the radius of the coated wire. Film-thickening in the presence of surfactants was also found in plate-coating applications \cite{Krechetnikov_pof_2005,Campana_pof_2010}, often referred to as the `Landau-Levich problem'. 

Other numerical works have focused on the effect of soluble surfactants on liquid displacement by a gas phase in the negligible inertia regime \cite{Severino_pof_2003,Ghadiali_jfm_2003}. Severino {\it et al.} \cite{Severino_pof_2003} reported film thickening in all cases, whereas Ghadiali and Gaver \cite{Ghadiali_jfm_2003} found that for bulk Peclet number $Pe_c>10$, where $Pe_c=U_bR/D_c$ with $D_c$ being the bulk diffusion coefficient, or low adsorption rates, film-thinning may occur. Further computational efforts by Johnson and Borhan \cite{Johnson_jciS} investigated the effect of soluble surfactants on bubble motion and concluded that at low surface coverage, reduced drop mobility is uniform, whereas at high concentration a stagnant bubble cap forms. More recently, Olgac and Muradoglu \cite{Olgac_ijmf_2013} performed an extensive study of the effect of a wide range of surfactant parameters on the film thickness of the bubbles.

The review of the literature above highlights the fact that the existing studies have primarily focused on the impact of surfactants in flow conditions where inertia is negligible. The aim of this work is to investigate the effect of surfactants on elongated gas bubbles propagating through capillary tubes, when inertial forces have a significant impact on the liquid film dynamics ($Re\gg 1$). In particular, this work will focus on the effect of surfactant addition on the undulatory structures observed at the bubble tail. The dynamics of the tail undulations plays a role for a number of engineering and scientific applications such as microchannel flow boiling \cite{Magnini_ijts_2016}, with specific emphasis on the potential occurrence of liquid film rupture and dryout \cite{Borhani_ijmf_2014}, and cleaning of microorganisms from the walls of confined microgeometries \cite{Khodaparast_est_2017}. In these applications, water and other low-viscosity refrigerant fluids are utilised, at high flow rates, such that the Reynolds number may exhibit values of $Re\sim 10^3$ even in sub-millimetric capillaries \cite{Khodaparast_mano_2015}.

We employ fully three-dimensional direct numerical simulations, using a hybrid interface-tracking/level-set method (also known as the Level Contour Reconstruction Method), proposed by Shin {\it et al.} \cite{Shin_jcp_2002,Shin_jcp_2005,Shin_jmst_2007,Shin_jmst_2017,Shin_jcp_2018}, where the unsteady dynamics of the free-interface is resolved explicitly. A comprehensive computational study is performed to assess the influence of a range of  dimensionless groups associated with the flow (e.g. Reynolds and capillary numbers) and the surfactants properties (e.g. Peclet number, elasticity number, Biot number, Damkohler number, and adsorption depth), on the bubble dynamics. The rest of this paper is organised as follows: in Sec. \ref{sec:form} the governing equations are presented, along with a description of the computational set-up, scaling, 
and validation of the numerical procedure. The main results and discussion are presented in Sec. \ref{sec:results}, where the overall effect of surfactant addition is discussed first, followed by a thorough parametric study. Finally, the main conclusions of this work are detailed in Sec. \ref{sec:conc}.
\section{Formulation and problem statement} \label{sec:form}
\subsection{Governing equations}
\begin{figure}
     \begin{subfigure}[b]{1\textwidth}
         
         \includegraphics[width=0.9\linewidth]{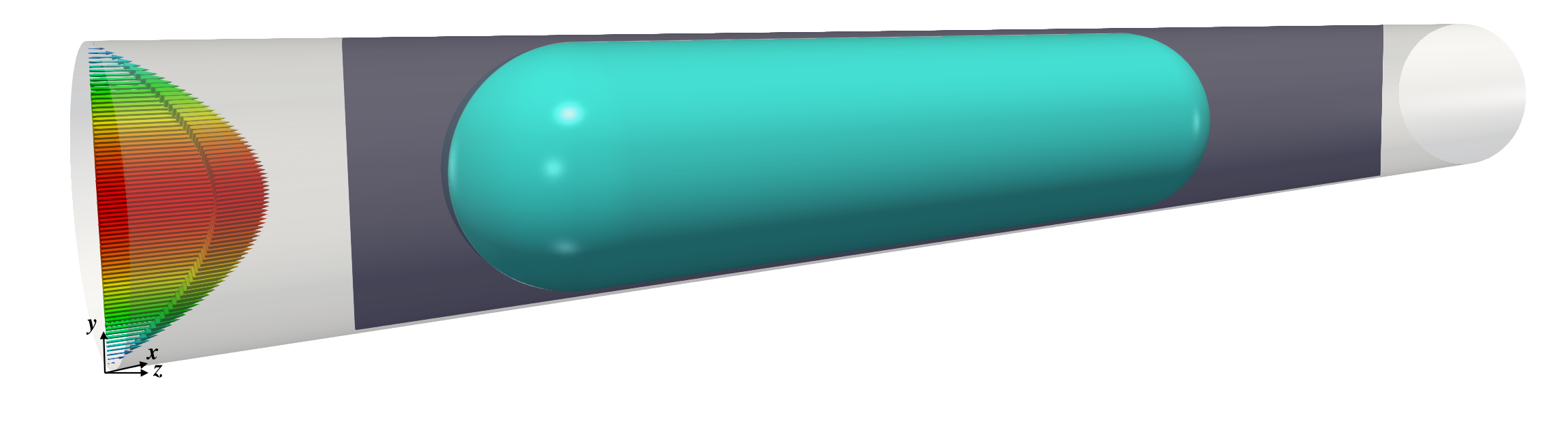}
         \caption{}
         \label{fig1_1}
     \end{subfigure}
          \hfill
        \begin{subfigure}[b]{1\textwidth}
         \includegraphics[width=0.8\linewidth]{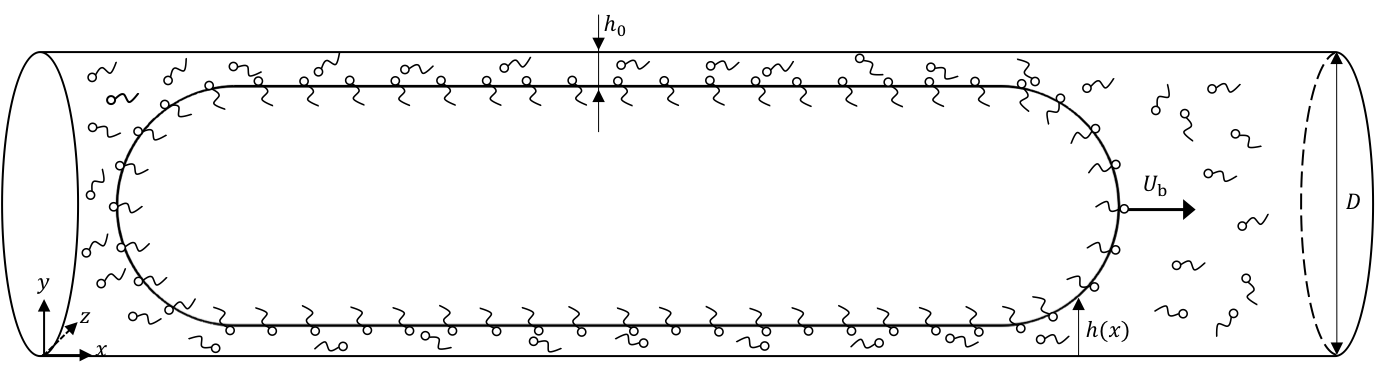}
         \caption{}
         \label{fig1_2}
     \end{subfigure}
        \caption{(a) Initial three-dimensional bubble shape and parabolic velocity profile imposed at the tube inlet. (b) Schematic representation of the problem in a vertical centreline  $(x,y)$ plane, where $D$ denotes the tube diameter, $U_b$ is the bubble-tip velocity, $h$ is the vertical distance of the liquid-gas interface from the $y=0$ axis, and $h_0$ is the uniform liquid film thickness.}
        \label{fig1}
     \centering
\end{figure}

In this section, the governing equations are presented in the context of the Level Contour Reconstruction Method (LCRM). The propagation of a gas bubble through a liquid-filled cylindrical tube of diameter $D$ is considered, as shown in Fig. \ref{fig1}. The gas and liquid are assumed to be immiscible, incompressible Newtonian fluids, and gravitational effects are neglected. The continuity and momentum equations are written in a three-dimensional Cartesian domain using a single-fluid formulation, respectively:
    \begin{equation}
    \nabla \cdot \textbf{u} =0, ~~~~~~~
%
    \label{momentum}
    \rho \left(\frac{\partial  \textbf{u}}{\partial t} + \textbf{u} \cdot \nabla \textbf{u}\right) =  - \nabla p +  \nabla \cdot \mu ({\nabla \textbf{u}}+{\nabla \textbf{u}}^T) + \int_{A}  
    \sigma  \kappa   \textbf{n} \delta   (\textbf{x}-\textbf{x}_f) dA 
    + \int_{A} 
    \nabla_s \sigma    \delta   (\textbf{x}-\textbf{x}_f)dA,~~~
    \end{equation} 
\noindent
where $t$, $\textbf{u}$, and $p$ denote  time, velocity, and pressure, and the density $\rho$ and viscosity $\mu$ are given by
    \begin{equation}
    \left.\begin{array}{c}
    \rho\left( \textbf{x},t\right)=\rho_b + \left(\rho_l -\rho_b\right) I\left( \textbf{x},t\right),\\
    \mu\left( \textbf{x},t\right)=\mu_b+ \left(\mu_l -\mu_b\right) I\left( \textbf{x},t\right).\\
    \end{array}\right.
    \end{equation}
Here, $I\left( \textbf{x},t\right)$ represents a smoothed Heaviside function, which is zero in the bubble (gas) phase and unity in the liquid phase, while the subscripts $l$ and $b$ designate the individual liquid and bubble phases, respectively.
The last two terms on the right-hand-side of Eq. (\ref{momentum}) represent the normal and tangential components of the surface tension force, respectively. The former is associated with the mean surface tension, $\sigma$, whereas the latter appears as a consequence of surface tension gradients, giving rise to Marangoni stresses; $\kappa$ denotes the interface curvature, $\nabla_s$ is the surface gradient operator, and $\textbf{n}$ is the outward-pointing unit normal to the interface. The three-dimensional Dirac delta function, $\delta   (\textbf{x}-\textbf{x}_f)$, vanishes everywhere except at the interface localised at $\textbf{x}=\textbf{x}_f$. 

The surfactant concentration on the interface, $\Gamma$, is governed by the following conservation equation
	\begin{equation}
	 \frac{\partial \Gamma}{\partial t}+\nabla_s \cdot (\Gamma\textbf{u}_t)=D_s \nabla^2_s \Gamma+ \dot{S_{\Gamma}},
	 \label{gamma}
	\end{equation}
where $\textbf{u}_t=(\textbf{u}_s\cdot\textbf{t})\textbf{t}$ is the tangential velocity vector in which $\textbf{u}_s$ is the surface velocity and ${\mathbf{t}}$ is the unit tangent to the interface. The diffusion of surfactant along the interface is accounted for in the first term on the right-hand-side, where $D_s$ is the surface diffusion coefficient. The sorptive flux, which characterises the exchange of surfactant species between the bulk and the interface, is given by the source term
	\begin{equation}
	\label{source}
	\dot{S_{\Gamma}}=k_a C_s (\Gamma_\infty-\Gamma)-k_d\Gamma,
	\end{equation}
where $k_a$	and $k_d$ are adsorption and desorption coefficients, respectively, $C_s$ is the concentration of surfactant in the bulk sub-phase, immediately adjacent to the interface, and $\Gamma_\infty$ is the interfacial surfactant concentration at saturation.
The transport of surfactant concentration $C$ in the bulk is governed by
    \begin{equation} 
    \frac{\partial C}{\partial t}+\textbf{u}\cdot\nabla C=D_c \nabla\cdot(\nabla C), 
	\end{equation}
where $D_c$ refers to the surfactant diffusivity in the bulk phase. The source term in Eqs. (\ref{gamma}) and (\ref{source}) can be related to the bulk concentration by
	\begin{equation}
	\label{bulksource}
	\textbf{n}\cdot\nabla C |_{interface}=-\frac{\dot{S_{\Gamma}}}{D_c},
	\end{equation}

The equation of state used in this work to describe the decrease of $\sigma$ with $\Gamma$ is given by the Langmuir relation \cite{Muradoglu_jcp_2014,Shin_jcp_2018}:
    \begin{equation} 
    \label{sigma}
    \sigma=\sigma_s+\Re T\Gamma_\infty \ln\left(1-\frac{\Gamma}{\Gamma_\infty}\right)=\sigma_s\left[1+\beta_s \ln\left(1-\frac{\Gamma}{\Gamma_\infty}\right)\right],
    \end{equation}
\noindent
where $\sigma_s$ is the surface tension of the `clean' interface, $\Re$ is the ideal gas constant, $T$ is temperature, and $\beta_s= \Re T\Gamma_\infty/\sigma_s$ is defined as the surfactant elasticity parameter. 

All variables are rendered dimensionless by using the following scalings:
\begin{equation}
\quad \tilde{\textbf{x}}=\frac{\textbf{x}}{D},
\quad \tilde{\textbf{u}}=\frac{\textbf{u}} {U}, 
\quad \tilde{t}=\frac{t}{D/U}, 
\quad \tilde{p}=\frac{p}{\rho_l U^2}, 
\quad \tilde{\sigma}=\frac{\sigma}{\sigma_s},
\quad \tilde{\Gamma}=\frac{\Gamma}{\Gamma_\infty},
\quad \tilde{C}=\frac{C}{C_{\infty}},
\quad \tilde{C_s}=\frac{C_s}{C_{\infty}},
\end{equation}
where the tildes designate dimensionless quantities. 
\noindent Here, the diameter $D$,  the average liquid velocity at the tube inlet $U$, and $C_\infty$, are used as the characteristic length, velocity, and bulk concentration scales. 
As a result of this scaling, Eqs. (\ref{momentum})-(\ref{sigma}) become:
\begin{equation}
 \nabla \cdot \tilde{\textbf{u}}=0, ~~~~
 \tilde{\rho}\left(\frac{\partial \tilde{\textbf{u}}}{\partial \tilde{t}}+\tilde{\textbf{u}} \cdot\nabla \tilde{\textbf{u}}\right)  = - \nabla \tilde{p}  +  \frac{1}{Re}\nabla\cdot  \left [ \tilde{\mu} (\nabla \tilde{\textbf{u}} +\nabla \tilde{\textbf{u}}^T) \right ] +
\frac{1}{Re  Ca} \int_{\tilde{A}} \left(
\tilde{\sigma} \tilde{\kappa} \textbf{n} 
 +  
 \nabla_s  \tilde{\sigma}  \right)\delta \left(\tilde{\textbf{x}} - \tilde{\textbf{x}}_{_f}  \right) d\tilde{A} , 
\end{equation}
    \begin{equation}
    \left.\begin{array}{c}
    \tilde{\rho}\left( \textbf{x},t\right)=\dfrac{\rho_b}{\rho_l} + \left(1 -\dfrac{\rho_b}{\rho_l}\right) I\left( \textbf{x},t\right),\\
    \tilde{\mu}\left( \textbf{x},t\right)=\dfrac{\mu_b}{\mu_l}+ \left(1 -\dfrac{\mu_b}{\mu_l}\right) I\left( \textbf{x},t\right),\\
    \end{array}\right.
    \end{equation}

\begin{equation} 
\frac{\partial \tilde{C}} {\partial \tilde{t}}+\tilde{\textbf{u}}\cdot \nabla \tilde{C}= \frac{1}{Pe_c} \nabla\cdot(\nabla \tilde{C}),
\end{equation}
 \begin{equation} 
 \frac{\partial \tilde{\Gamma}}{\partial \tilde{t}}+\nabla_s \cdot (\tilde{\Gamma}\tilde{\textbf{u}_t})=\frac{1}{Pe_s} \nabla^2_s \tilde{\Gamma}+ Bi \left ( k  \tilde{C_s} (1-\tilde{\Gamma})- \tilde{\Gamma}  \right ),
 \end{equation}
\begin{equation}
\label{bulksource_nd}
\textbf{n}\cdot\nabla \tilde{C} |_{interface}=-Pe_c Da Bi \left ( k  \tilde{C_s} (1-\tilde{\Gamma})- \tilde{\Gamma}  \right ),
\end{equation}
\begin{equation} 
\tilde{\sigma}=1 + \beta_s \ln{\left(1 -\tilde{\Gamma}\right)}.
\label{marangoni_eq}
\end{equation}
The dimensionless parameters appearing in these equations are given by

        \begin{equation}\label{dimless}
        Re=\frac{\rho_lUD}{\mu_l};~Ca=\frac{\mu_lU}{\sigma_s};~
        Pe_c=\frac{UD}{D_c};~Pe_s=\frac{UD}{D_s};~ 
        Bi=\frac{k_d D}{U};~Da=\frac{\Gamma_\infty}{D C_\infty};
        ~k=\frac{k_a C_\infty}{k_d},
        \end{equation}
\noindent
where $Ca$ and $Re$ are the liquid capillary and Reynolds numbers, and the density and viscosity ratios are represented by $\rho_l/\rho_b$ and $\mu_l/\mu_b$, respectively. The competition between convection and diffusion for the surfactant species at the interface and in the bulk is characterised by $Pe_s$ and $Pe_c$, respectively. Other surfactant-related parameters are the 
Biot number, $Bi$, Damkohler number, $Da$, 
and the adsorption depth, $k$. The bulk surfactant concentration, used as an initial condition, and kept constant throughout the simulation, is represented by $C_\infty$. 
The surface tension gradients give rise to the Marangoni stresses, which can be expressed in terms of $\Gamma$ by
\begin{equation}
\frac{1}{Re Ca} \nabla_s \tilde{\sigma}\cdot {\mathbf{t}}  \equiv \frac{\tilde{\tau}}{Re Ca}=-Ma\frac{1}{(1-\tilde{\Gamma})}\nabla_s\tilde{\Gamma}\cdot {\mathbf{t}},  
\end{equation}
where $Ma\equiv \beta_s/Re \,Ca= \Re T\Gamma_\infty/\rho_l U^2 D$ is a Marangoni parameter.

\subsection{Problem statement and validation}

The flow domain is a tube of circular cross-section of diameter $D$ and length $27.8 D$, which is modeled with a three-dimensional geometry. The walls of the tube are constructed via a module that defines solid objects by means of a signed distance function, an approach that was previously adopted for more complex geometries when using the present numerical solver \cite{Kahouadji_mano_2018,Russell_ces_2019}.
The elongated bubble is initially located at the beginning of the channel and its shape is initialised using a cylindrical body with a cross-sectional diameter $D_b=0.94 D$, and two hemispherical caps at its two ends, as indicated in Fig. \ref{fig1_1}. We adopt a reference frame where $x$ represents the streamwise coordinate, $y$ the vertical coordinate with $y=0$ being the bottom line of the tube, and $z$ the horizontal coordinate. The length of the bubble, $L_b$, is one of the parameters varied in this work. The flow is initiated by imposing a fully-developed parabolic velocity profile at the inlet (i.e. $x=0$). A no-slip boundary condition is imposed on the channel wall. The channel length is set to be sufficiently long to allow for the development of steady-state motion of the bubble.

The effect of the governing dimensionless groups in Eq. (\ref{dimless}) is explored through a systematic parametric study throughout which the density and viscosity ratios are kept constant at $\rho_l/\rho_b=1000$ and $\mu_l/\mu_b=100$, respectively, representing the values associated with an air-water system. A `base' case is defined, characterised by the following 
dimensionless parameters listed in Eq.~\eqref{dimless}: 
$Re=443$, $Ca=0.0089$, 
$Pe_c=100$, $Pe_s=100$, $\beta_s=0.5$, $Bi=1$, $Da=0.1$, $k=1$, and $Ma=0.13$; the initial dimensionless length of the bubble for the base case is kept as $\tilde{L}_b\equiv L_b/D=5$. 
At equilibrium, 
adsorption and desorption are equal and 
the source term in Eq. (\ref{source}) is equal to zero. Assuming that the surfactant concentration near the interface is equal to the surfactant concentration in the bulk $C_s=C_\infty$, the following relationship for $\Gamma_{eq}$ can be obtained
    \begin{equation}
        \frac{\Gamma_{eq}}{\Gamma_\infty}=\frac{k}{k+1}.
    \end{equation}
\noindent
For the base case (e.g. $k=1$), this relationship becomes $\Gamma_{eq}=0.5\Gamma_\infty$. The initial interfacial surfactant concentration is specified to be $50\%$ of the equilibrium concentration. 

The computations are assumed to reach steady-state when the change in total bubble concentration is below $0.1\%$ for one period of dimensionless time as suggested by Olgac and Muradoglu \cite{Olgac_ijmf_2013}. In addition, the steady propagation of the bubble is also monitored. The `clean' case is presumed to be at steady-state when the change in velocity of the bubble front and back menisci is less than $0.1\%$. 
As suggested by Ratulowski and Chang \cite{Ratulowski_jfm_1990}, a balance between adsorption and interfacial convection is achieved if the Stanton number, $St=k_a \Gamma_\infty/U$, is $St \approx O (Ca^{1/3})$, which is the case for the selected base case parameters. Following the guidelines by Ratulowski and Chang \cite{Ratulowski_jfm_1990}, the selected bulk Peclet number represents a system where both convective and diffusive transport ($Pe_c \approx O(Ca^{-2/3})$) parts play a role in the dynamics and both need to be resolved.

\begin{figure}
     \begin{subfigure}[b]{0.49\textwidth}
         \includegraphics[width=1\linewidth]{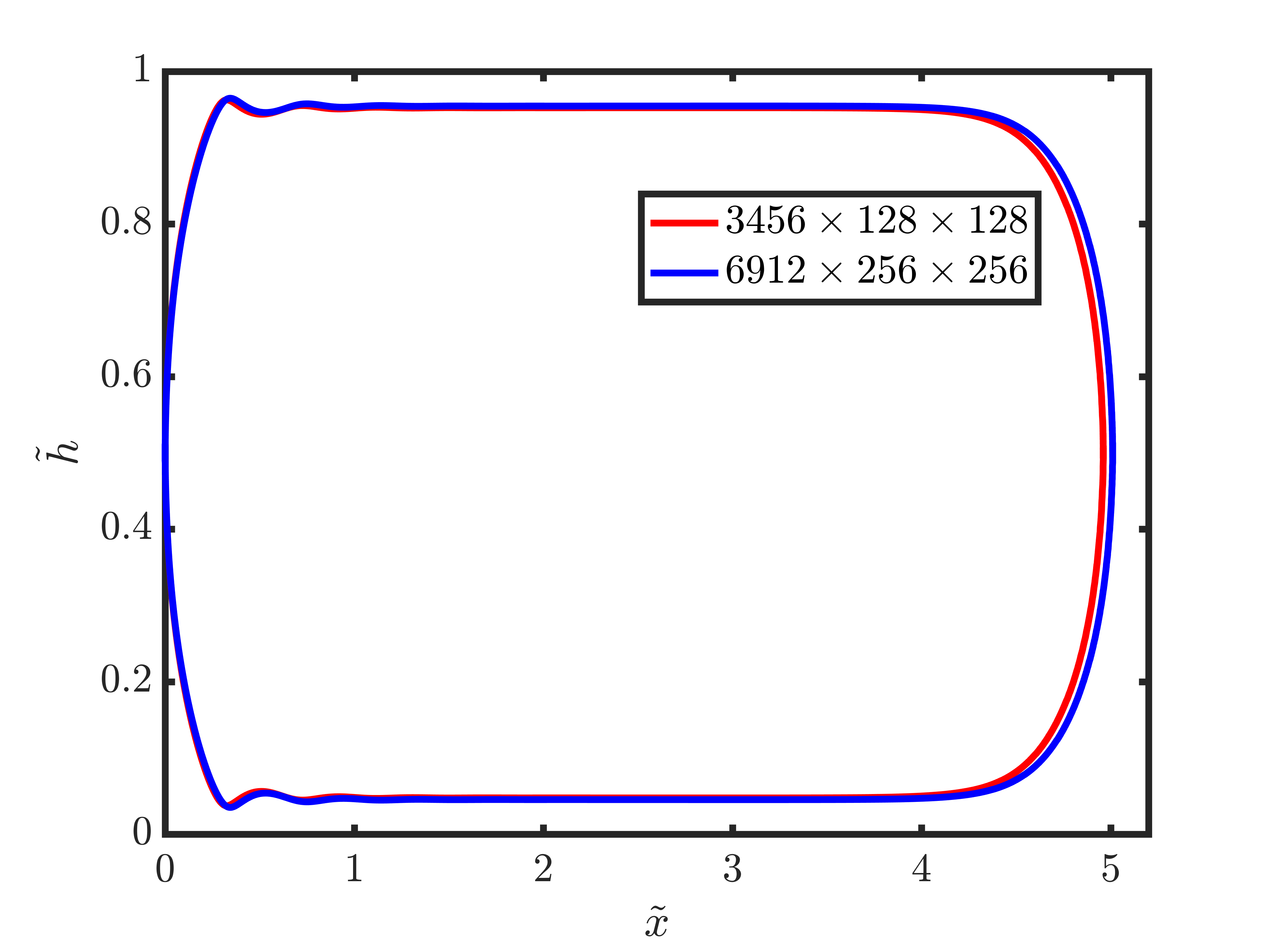}
         \caption{}
         \label{fig2_2}
     \end{subfigure}
          \hfill
        \begin{subfigure}[b]{0.49\textwidth}         
         \includegraphics[width=1\linewidth]{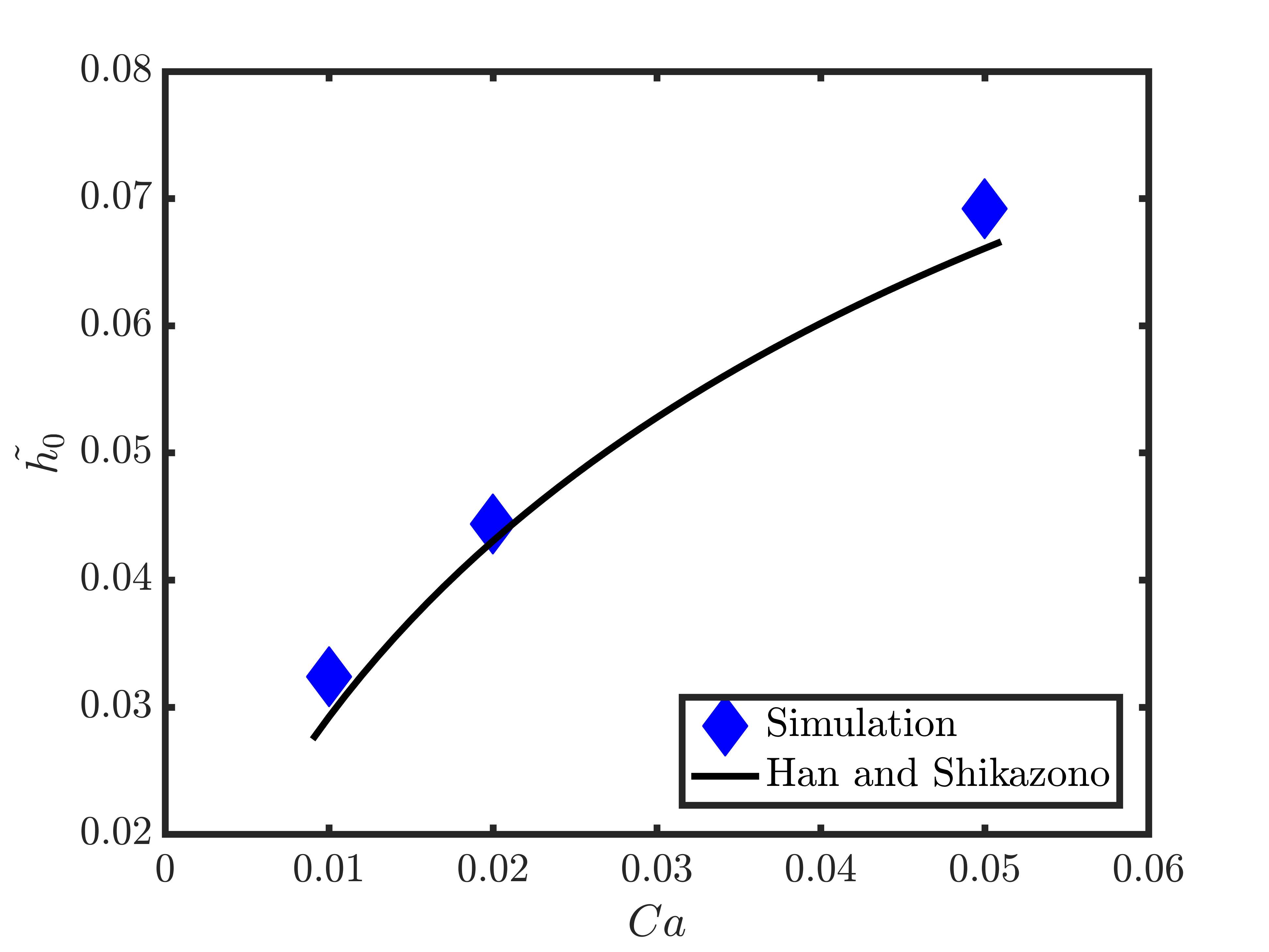}
         \caption{}
         \label{fig2_1}
     \end{subfigure}
        \caption{(a) Mesh dependence study for a `clean' bubble with reference parameters $Re=443$ and $Ca=0.0089$; (b) comparison of the experimental correlation of Han and Shikazono \cite{Han_ijhff_2009} with the simulation results from this study, for a varying capillary number and fixed $Re=443$.}
        \label{fig2}
\end{figure}

The computational work in this study employs a three-dimensional uniform Cartesian grid. The grid dependence analysis (see Fig. \ref{fig2_2}), performed for the base case surfactant-free bubble parameters, shows that doubling the cell count in each direction has a negligible effect on the bubble shape. All subsequent simulations are performed using the coarser grid of 56.6 million cells (i.e. $3456 \times 128 \times 128$). Validation is performed against the empirical correlation for film thickness presented by Han and Shikazono \cite{Han_ijhff_2009} and given by Eq. (\ref{HS}). The film thickness in the computational results is evaluated at the midpoint between the bubble nose and tail, where the liquid film is uniform. At this location, the value of the film thickness is representative of an average value for the bubble. The numerical results are found to be in good agreement with the experiment, with a maximum $10\%$ deviation that is within the uncertainty of the experimental correlation \cite{Han_ijhff_2009}. 

\section{Results and discussion} \label{sec:results}
\subsection{Effect of Marangoni stresses} 
\label{sec:overall}
In this section, the effect of key flow and surfactant dimensionless groups on the bubble dynamics is studied. Firstly, the effect of the base case surfactant parameters is considered. As described previously, the addition of surface active species acts to reduce the surface tension and gives rise to surface tension gradients. In order to isolate the effect of the Marangoni stresses, a simulation is performed where the surface tension, $\sigma$, is set  
equal to the steady-state average surface tension for the surfactant base case. As $\sigma=\sigma(\Gamma)$, the average interfacial concentration is evaluated  
and then used to calculate the effective surface tension, $\sigma_{eff}$, given by Eq. (\ref{sigma}). To distinguish between the cases, a capillary number based on the effective surface tension is defined for this section, $Ca_{eff}=\mu_lU/\sigma_{eff}$. This effective capillary number for the surfactant base case is  
$Ca_{eff}=0.0094$. Although the change in capillary number is not expected to influence the bubble dynamics significantly, the $Ca_{eff}=0.0094$ case allows us to separate the effect of surfactants on reducing the mean surface tension, from that associated with the formation of Maragoni stresses due to surface tension gradients. This case will be referred to below as the `no-Marangoni' case, or $\tilde{\tau}=0$. 

\begin{figure}[htbp!]     
     \begin{subfigure}[b]{\textwidth}
         \includegraphics[width=0.8\linewidth]{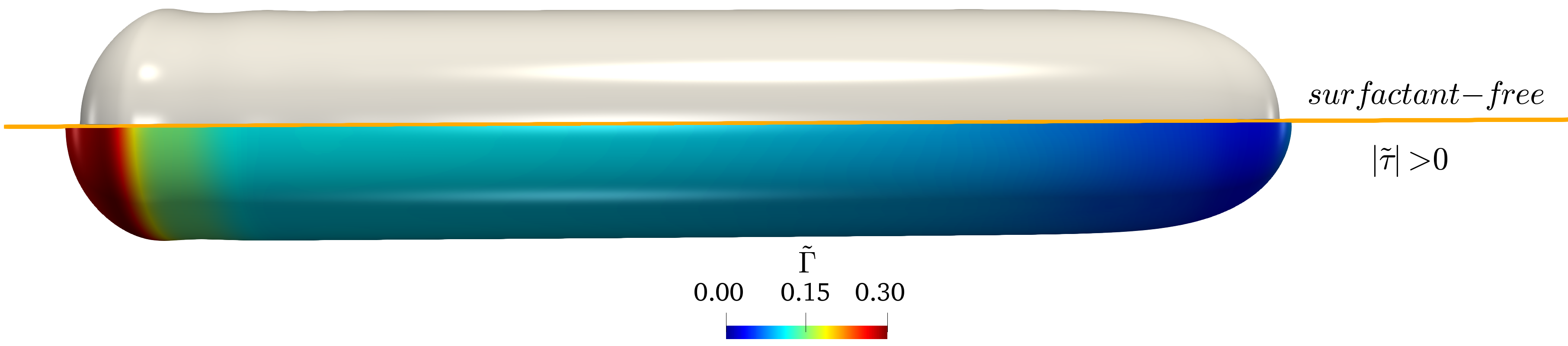}
         \caption{}
         \label{fig3_1}
     \end{subfigure}
     \hfill
     \begin{subfigure}[b]{1\textwidth}
         
         \includegraphics[width=1\linewidth]{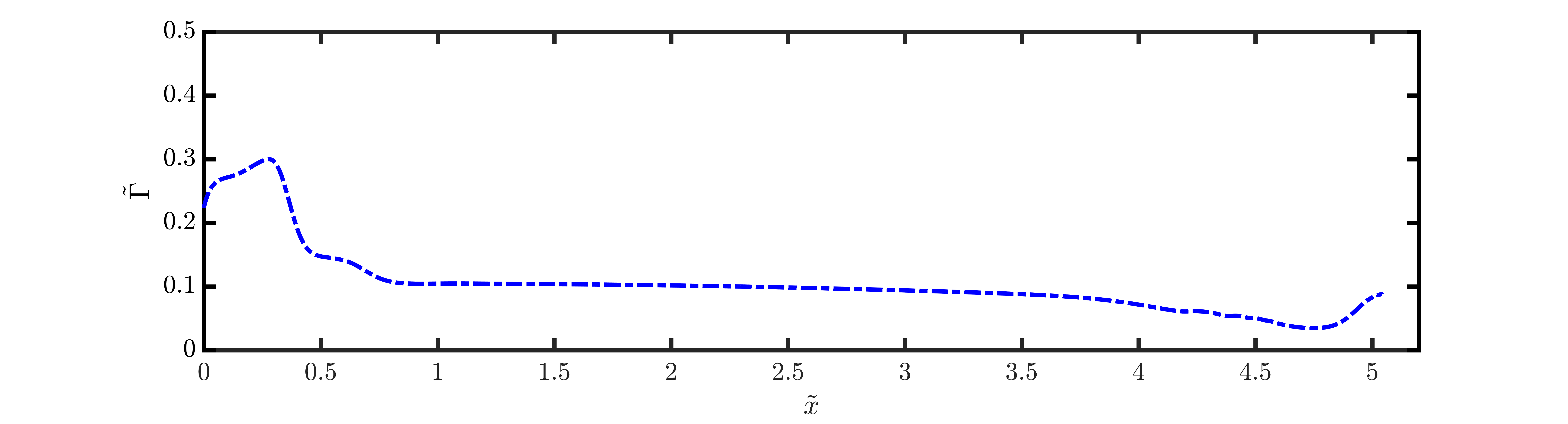}
         \caption{}
         \label{fig3_2}
     \end{subfigure}
          \begin{subfigure}[b]{1\textwidth}
         
         \includegraphics[width=1\linewidth]{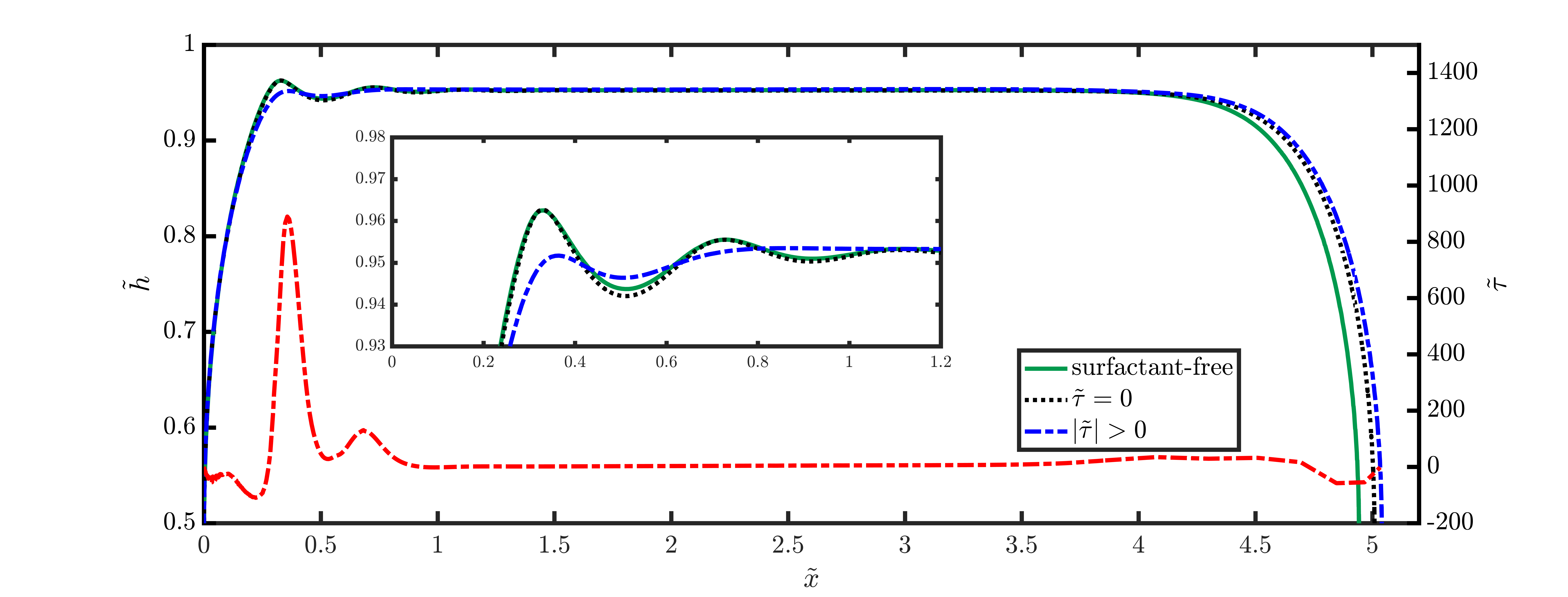}
         \caption{}
         \label{fig3_3}
     \end{subfigure}
        \caption{Effect of surfactant on the flow characteristics; (a) three-dimensional representation of the bubble shape for the surfactant-free (top) and surfactant-laden (bottom) cases, with the colour indicating the magnitude of surfactant interfacial concentration, $\tilde{\Gamma}$; (b) variation of $\tilde{\Gamma}$ along $\tilde{x}$; (c) two-dimensional projection (in the $z=0$ plane) of the bubble shape for the surfactant-free (solid line), and  surfactant-laden cases in the presence (dashed) and absence (dotted) of Marangoni stresses. Also shown in (c), as a red dashed line, is the $\tilde{x}$ variation of the Marangoni stresses, $\tilde{\tau}$. The capillary numbers for the surfactant-free and surfactant-laden cases are 0.0089 and 0.0094, respectively, while the rest of the parameters are $Re=443$, $Pe_c=Pe_s=100$, $Da=0.1$, $k=1$, $Bi=1$, $\beta_s=0.5$, and $Ma=0.13$. }
        \label{fig:clean_surf}
\end{figure}

The effect of the base case surfactant parameters in comparison to the `clean' and `no-Marangoni' cases is presented in Fig. \ref{fig:clean_surf}. In Fig. \ref{fig3_1}, it is seen that
the surfactant is swept to the back of the bubble by the flow and accumulates in that region as also shown in Fig. \ref{fig3_2}, which illustrates the variation of the interfacial concentration, $\tilde{\Gamma}$, along $\tilde{x}$ for the Marangoni-supported case. The $\tilde{\Gamma}$ profile exhibits an increase to a peak value, which is spatially-coincident with the peak of the interfacial oscillation at the bubble tail shown in Fig. \ref{fig3_3}. The concentration then decreases via mild undulations towards an essentially constant value, which extends over a substantial fraction of the bubble; these $\tilde{\Gamma}$ variations are mirrored by similar characteristics in the bubble shape where the constant $\tilde{\Gamma}$ region coincides with that of the film of uniform thickness that separates the bubble from the wall. The concentration $\tilde{\Gamma}$ then undergoes a decrease followed by an increase in response of a stagnation point located near the bubble tip, as it will be revealed below by the analysis of the fluid flow.
The average film thickness for the surfactant base case is seen to decrease when compared to the `clean' and `no-Marangoni' cases by $2.5\%$ and $3.6\%$, respectively; similar observations were made by Ghadiali and Gaver \cite{Ghadiali_jfm_2003}. 

The non-uniform distribution in $\tilde{\Gamma}$ near the two extremes of the domain gives rise to concentration gradients and Marangoni stresses whose spatial variation is also shown in Fig. \ref{fig3_3}. A close comparison of the tail dynamics associated with the `clean' and `no-Marangoni' cases shown in the inset of Fig. \ref{fig3_3} reveals that these dynamics are essentially independent of the presence of surfactant; in contrast, the oscillations in the Marangoni-supported case are damped significantly thereby illustrating that the damping is Marangoni-driven. This is explained further via inspection of the surfactant distribution near the bubble tail, which induces Marangoni stresses that drive flow away from the peak of the interfacial oscillation, decreasing its amplitude. The Marangoni stresses are predominantly positive in this region and thus they act to retard the flow and `rigidify' the interface. As the right domain boundary is approached, the magnitude of the Marangoni stresses decreases considerably though they remain positive-valued apart from a narrow region in which they are negative in response to surfactant accumulation at the bubble nose.
The rigidifying effect of the Marangoni stress at the front and the back of the bubble reduces the speed of the bubble by approximately $5\%$, an effect previously observed by Borhan and Mao \cite{Borhan_PofA_1992}.

\begin{figure}
        \begin{subfigure}[b]{0.99\textwidth}
         \includegraphics[width=1\linewidth]{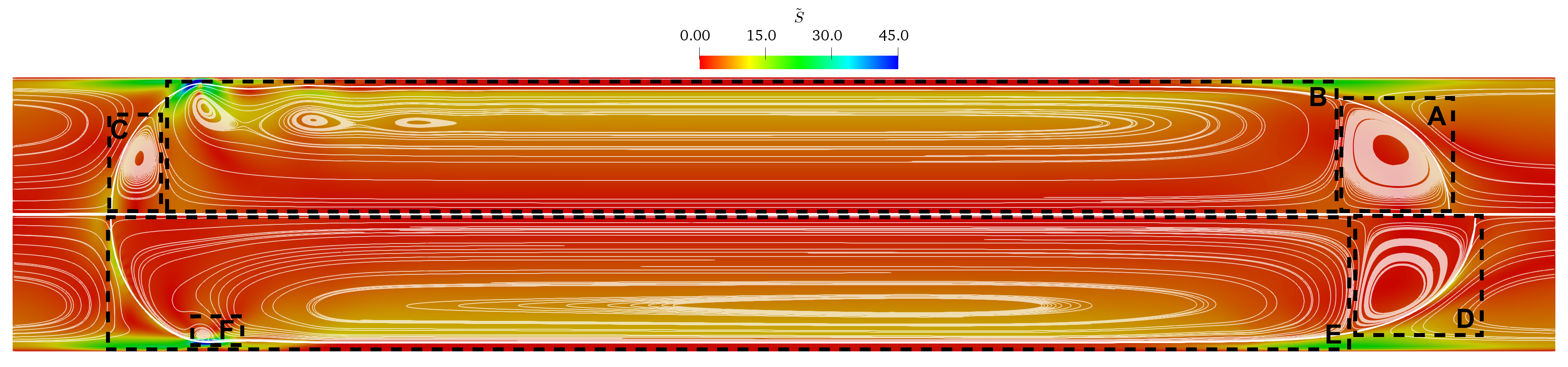}
         \caption{}
         \label{fig4_1}
     \end{subfigure}
     \hfill
     \begin{subfigure}[b]{0.99\textwidth}
         \includegraphics[width=1\linewidth]{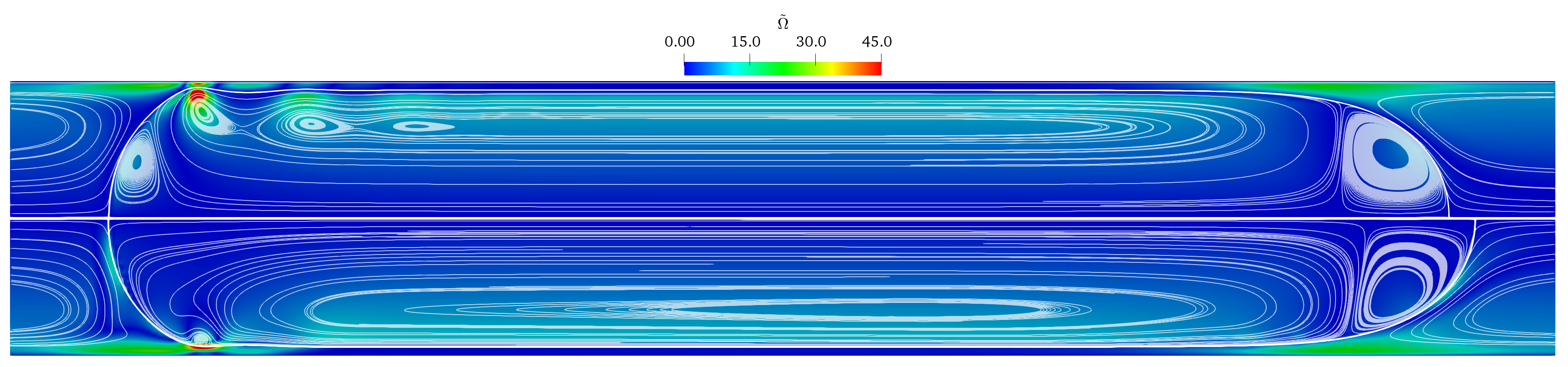}
         \caption{}
         \label{fig4_2}
     \end{subfigure}
        \caption{Effect of surfactant on the magnitude of the dimensionless strain rate $\tilde{S}$, (a), and vorticity $\tilde{\Omega}$, (b), for the surfactant-free (top) and Marangoni-supported surfactant-laden (bottom) cases for the same parameters as in Fig. \ref{fig:clean_surf}; here, $S$ and $\Omega$ are scaled on $D/U$. The streamlines are drawn in the bubble-tip reference frame. The vortical structures identified in zones `A', `C', and `E', and B', `D', and `F' rotate in a clockwise and counter-clockwise direction, respectively.}
        \label{fig:streamlines}
\end{figure}

In Fig. \ref{fig:streamlines}, we show the effect of surfactant on the strain rate and vorticity for the same parameter values as those used to generate Fig. \ref{fig:clean_surf}. A comparison of the strain rate and vorticity patterns associated with the `clean' and surfactant-laden, Marangoni-supported cases reveals some similarities in terms of the counter-rotating vortical structures in regions `A' and `B', and `D' and `E', in the former and latter cases, respectively. Note the stagnation point at the boundary between the regions 'E' and 'D' for the surfactant-laden case; here, the streamlines diverge driving the surfactant away from the stagnation point, which yields the local minimum of $\tilde{\Gamma}$ observed in Fig.~\ref{fig3_2}. It can also be seen that Marangoni-driven damping of the tail oscillations leads to suppression of the vortical structures in regions `B' and `C' that are a feature of the back of the `clean' bubble case. Vortex `F' in Fig. \ref{fig:streamlines} is related to the change in the sign of the Marangoni stresses near the back of the surfactant-laden bubble discussed above. 

\begin{figure}[htbp!] 
     \centering

     \begin{subfigure}[b]{\textwidth}
         \centering
         \includegraphics[width=1\linewidth]{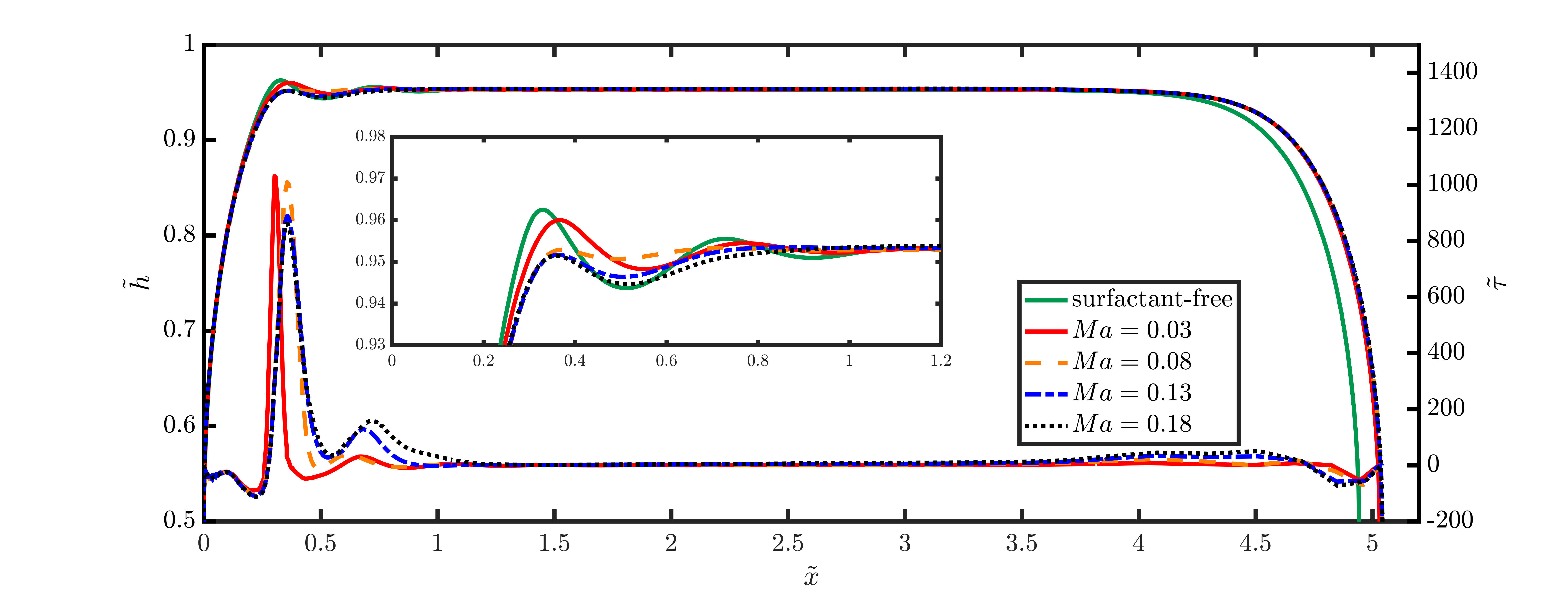}
         \caption{}
         \label{fig7_1}
     \end{subfigure}
     \hfill
     \begin{subfigure}[b]{\textwidth}
         \centering
         \includegraphics[width=1\linewidth]{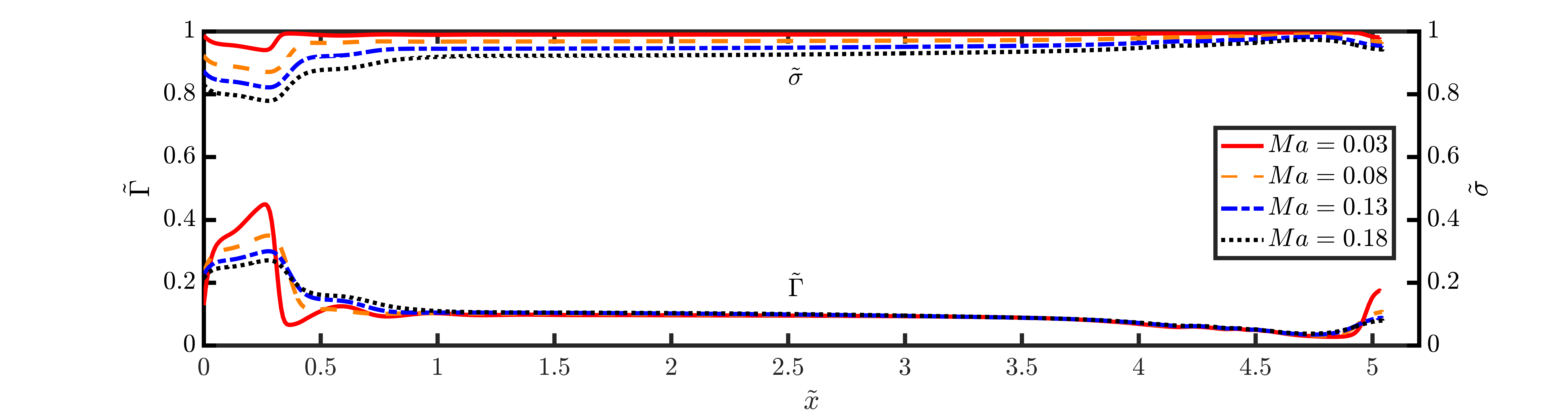}
         \caption{}
         \label{fig7_2}
     \end{subfigure}
     \hfill
    \begin{subfigure}[b]{\textwidth}
         \centering
         \includegraphics[width=1\linewidth]{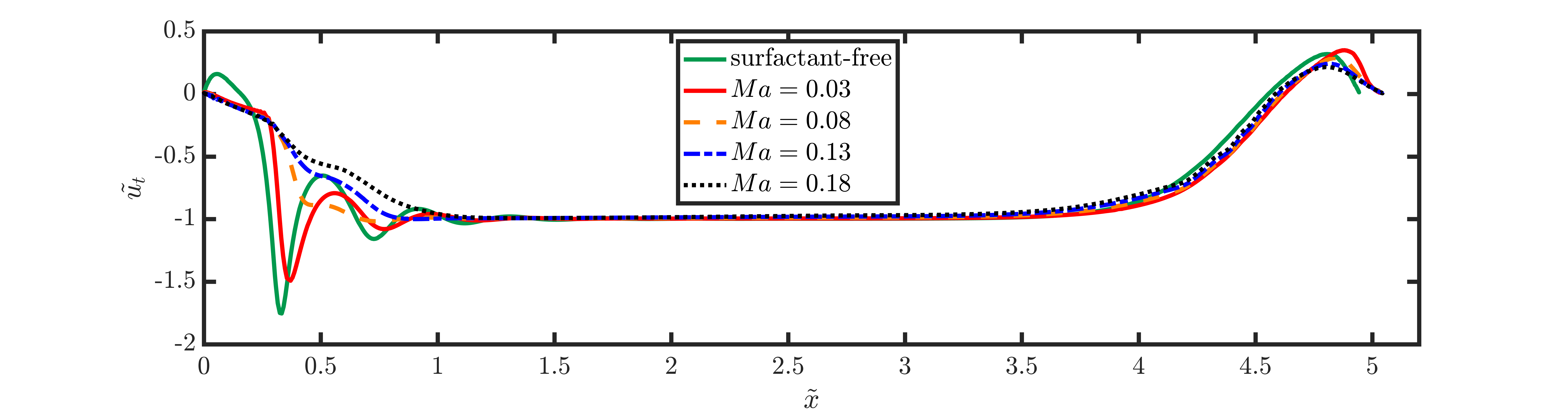}
         \caption{}
         \label{fig7_3}
     \end{subfigure}
     \hfill
        \caption{Effect of varying $Ma$ on the steady spatial distribution of the Marangoni stresses and two-dimensional projection (in the $z=0$ plane) of the bubble shape, (a), the interfacial surfactant concentration and resulting surface tension, (b), and the streamwise component of the interfacial velocity in the frame-of-reference of the bubble tip, (c), where $\tilde{u}_t=(u_t-U_b)/U_b$; here, $Ca=0.0089$ and the rest of the parameters remain unchanged from Fig. \ref{fig:clean_surf}.
        }
        \label{fig:beta}
\end{figure}

Next, we investigate the effect of altering the Marangoni parameter, $Ma$, on the flow profiles with $Ca=0.0089$ and $Re=443$ with all other parameters remaining unchanged from Fig. \ref{fig:clean_surf}. In Fig. \ref{fig:beta}, it is seen that increasing $Ma$ leads to more effective suppression of the bubble tail interfacial and $\tilde{u}_t$ oscillations, as shown in Fig. \ref{fig7_1} and \ref{fig7_3};
the increase in $Ma$ also results in a slight elongation of the bubble. The reduced interface mobility resulting from the rise in $Ma$ results in more uniform $\tilde{\Gamma}$ distributions, 
as can be seen in Fig. \ref{fig7_2}, and, therefore, the weakest $\tilde{\Gamma}$ gradients, and hence smallest steady Marangoni stresses.
As a result, the largest Marangoni stresses are those observed at the bubble tail for the lowest finite $Ma$ studied, as was also reported by Olgac and Muradoglu \cite{Olgac_ijmf_2013}. 

\subsection{Effect of $Ca$ and $Re$}
\begin{figure}[htbp!]  
     \centering
     \begin{subfigure}[b]{\textwidth}
         \centering
         \includegraphics[width=1\linewidth]{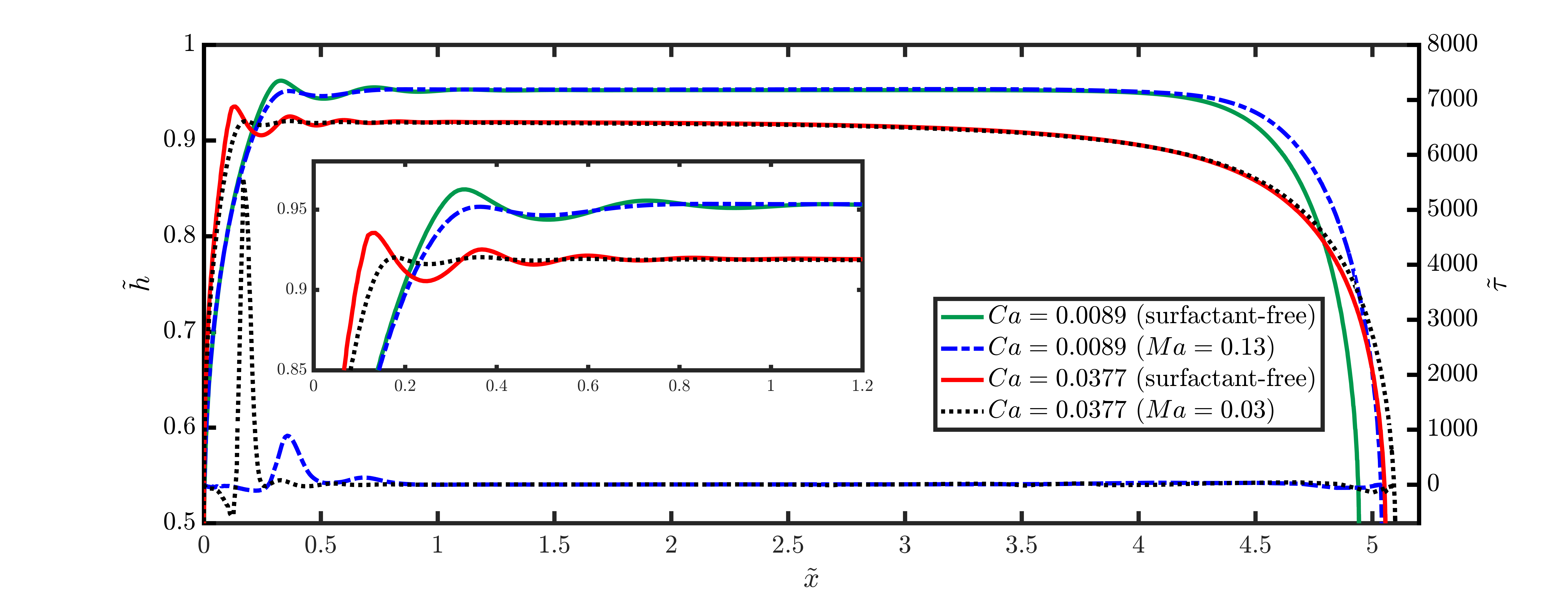}
         \caption{}
         \label{fig5_1}
     \end{subfigure}
     \hfill
     \begin{subfigure}[b]{\textwidth}
         \centering
         \includegraphics[width=1\linewidth]{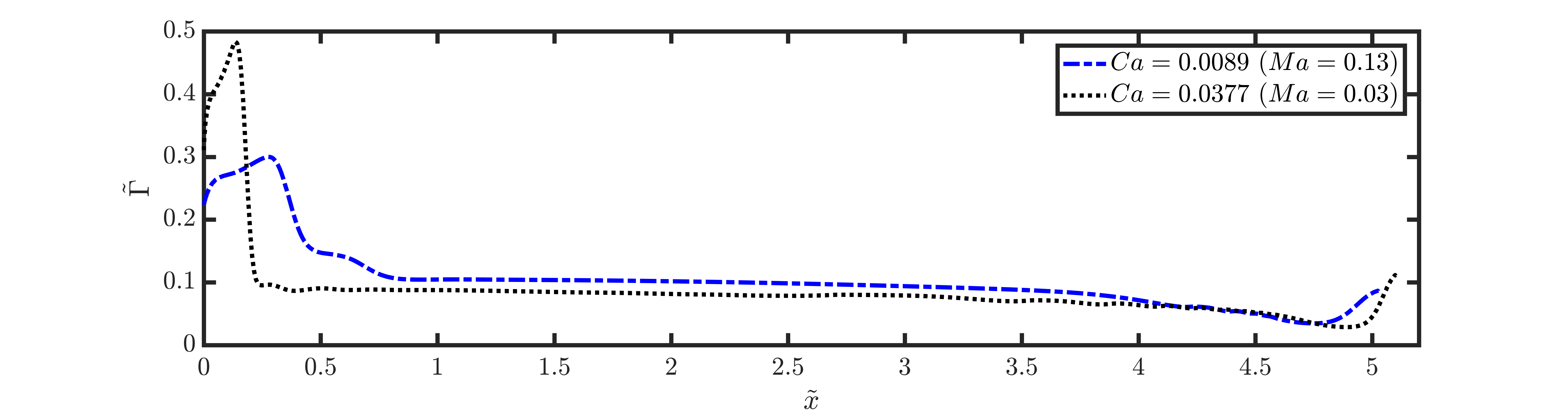}
         \caption{}
         \label{fig5_2}
     \end{subfigure}
     \hfill
     \begin{subfigure}[b]{\textwidth}
         \centering
         \includegraphics[width=1\linewidth]{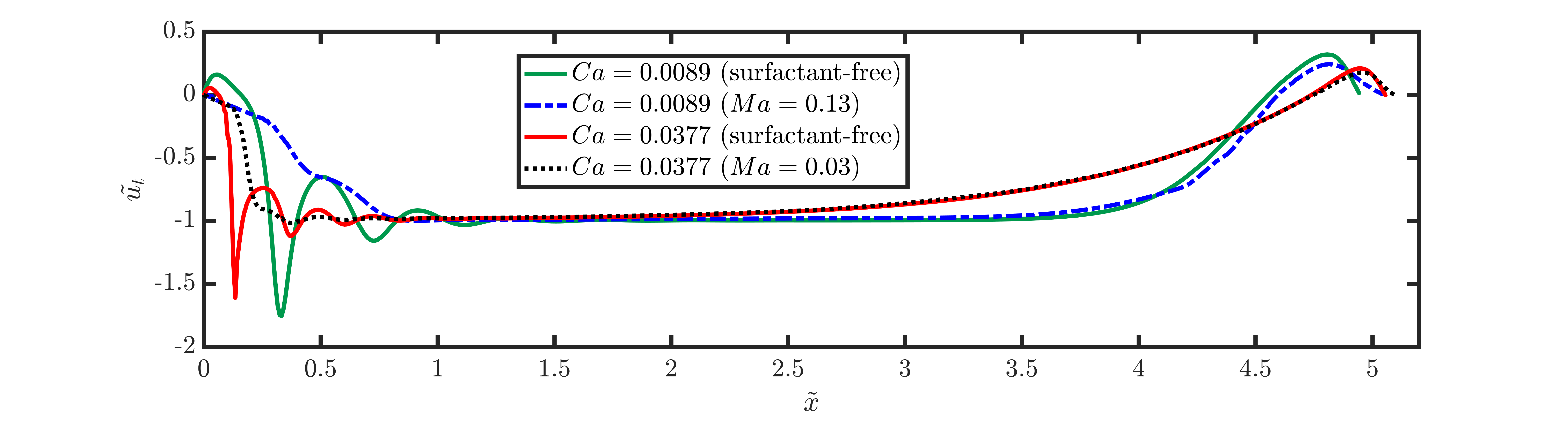}
         \caption{}
         \label{fig5_3}
     \end{subfigure}
        \caption{Effect of varying $Ca$ on the steady spatial distribution of the Marangoni stresses and two-dimensional projection (in the $z=0$ plane) of the bubble shape, (a), the interfacial surfactant concentration, (b), and the streamwise component of the interfacial velocity in the frame-of-reference of the bubble tip, (c), where $\tilde{u}_t=(u_t-U_b)/U_b$; the rest of the parameters remain unchanged from Fig. \ref{fig:clean_surf}. 
        }
        \label{fig:Ca}
\end{figure}

This section focuses on the effect of varying the Reynolds and capillary numbers in the presence of surfactant, where the base case surfactant parameters remain unchanged from $Pe_c=Pe_s=100$, $Da=0.1$ , $k=1$, $Bi=1$, $\beta_s=0.5$, and $Ma=0.13$. The investigation begins with variation of the capillary number as shown in Fig. \ref{fig:Ca} generated for $Ca=0.0089$ and $Ca=0.0377$ for both surfactant-free and surfactant-laden bubbles, with $Re=443$. 
Inspection of Fig. \ref{fig5_1} reveals that an increase in $Ca$ results in film thickening, while the amplitude of the interfacial undulations near the back of the bubble increases and their wavelength decreases with $Ca$, as also previously observed by Magnini {\it et al.} \cite{Magnini_prf_2017}.  
It is also seen clearly that the addition of surfactant
dampens these oscillations for both capillary numbers, as demonstrated in the inset of Fig. \ref{fig5_1}. This is due to the accumulation of surfactant at the bubble rear, depicted in Fig. \ref{fig5_2}, which leads to the formation of large Marangoni stresses in this region, as discussed in Sec. \ref{sec:overall}, whose magnitude increases with $Ca$ (see Fig. \ref{fig5_1}).

It is also instructive to examine the variation of the interfacial tangential velocity component in the bubble-tip reference frame $\tilde{u}_t\equiv (u_t-U_b)/U_b$, where $U_b$ is the bubble tip speed, along $\tilde{x}$, $\tilde{x}=x/D$, shown in Fig. \ref{fig5_3}. For all the cases considered, $\tilde{u}_t=0$ at the bubble tip due to the chosen moving frame-of-reference; $\tilde{u}_t$ then becomes positive-valued behind the tip before decreasing through zero, which coincides with the location of the stagnation point that separates the counter-rotating vortices at the bubble front discussed above in connection with Fig. \ref{fig:streamlines}. The tangential velocity assumes a value of $\tilde{u}_t=-1$, indicating a free-slip interface, over a significant proportion of the flat film region of the bubble before reaching $\tilde{u}_t=0$ at the bubble rear through oscillations that are damped severely in the surfactant-laden case, due to the rigidifying effect of the surfactant-induced Marangoni stresses. 

\begin{figure}[htbp!]  
     \centering
     \begin{subfigure}[b]{\textwidth}
         \centering
         \includegraphics[width=1\linewidth]{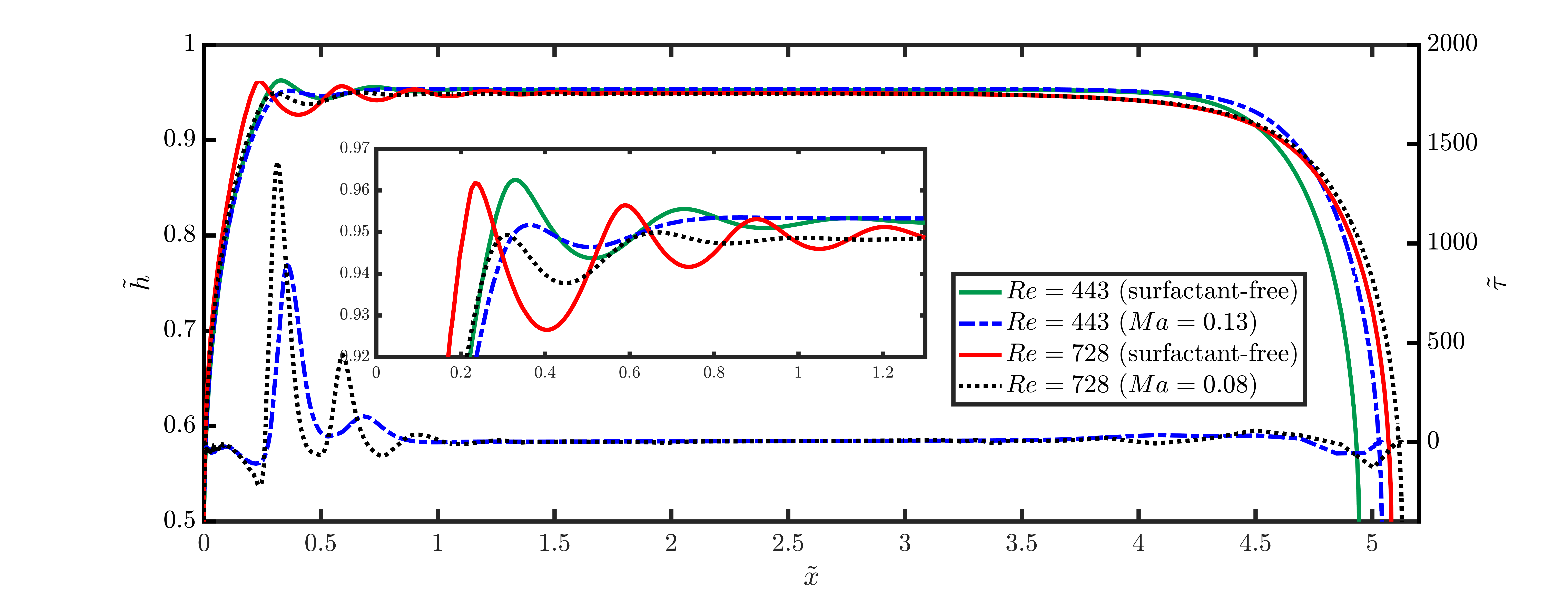}
         \caption{}
         \label{fig6_1}
     \end{subfigure}
     \hfill
     \begin{subfigure}[b]{\textwidth}
         \centering
         \includegraphics[width=1\linewidth]{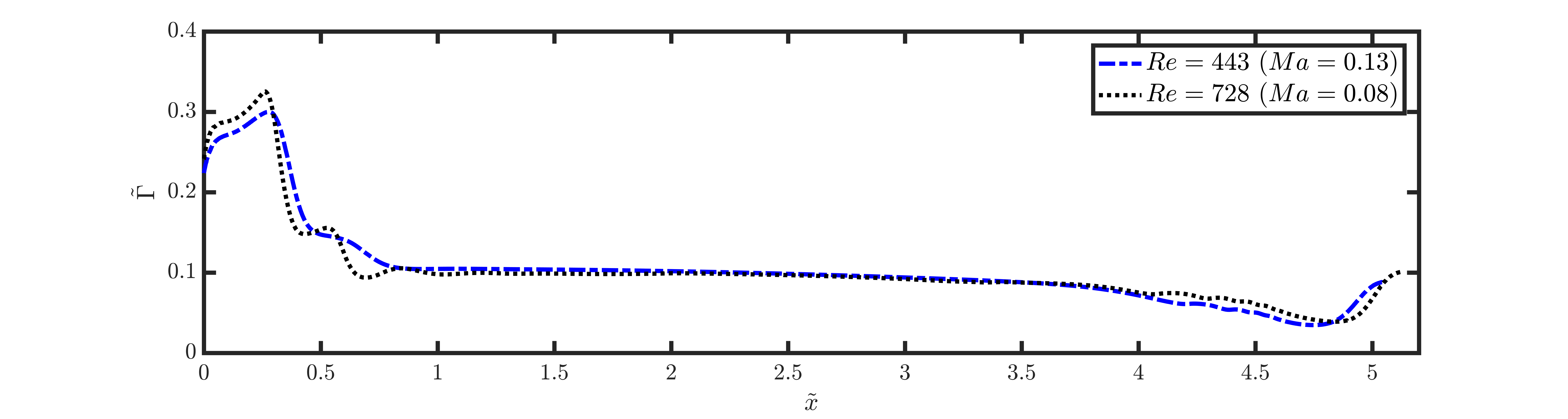}
         \caption{}
         \label{fig6_2}
     \end{subfigure}
     \hfill
     \begin{subfigure}[b]{\textwidth}
         \centering
         \includegraphics[width=1\linewidth]{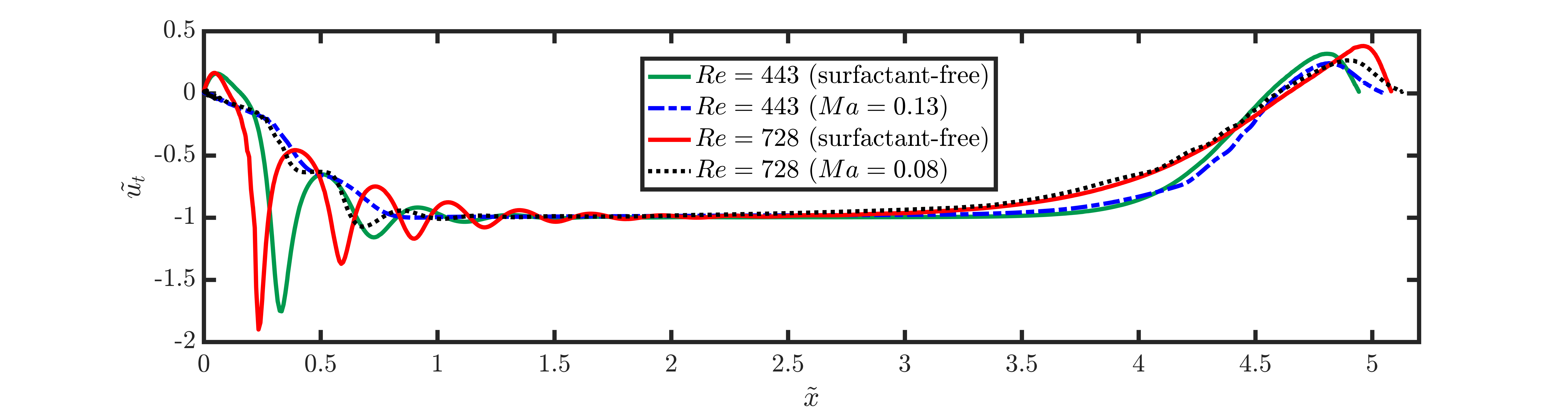}
         \caption{}
         \label{fig6_3}
     \end{subfigure}
        \caption{Effect of varying $Re$ on the steady spatial distribution of the Marangoni stresses and two-dimensional projection (in the $z=0$ plane) of the bubble shape, (a), the interfacial surfactant concentration, (b), and the streamwise component of the interfacial velocity in the frame-of-reference of the bubble tip, (c), where $\tilde{u}_t=(u_t-U_b)/U_b$; here, $Ca=0.0089$ and the rest of the parameters remaining unchanged from Fig. \ref{fig:clean_surf}. 
        }
        \label{fig:Re}
\end{figure}

In Fig. \ref{fig:Re}, we study the effect of increasing inertia on the bubble dynamics by raising $Re$ from $Re=443$ to $Re=728$ with  $Ca=0.0089$ and the rest of the parameters remaining unaltered from Fig. \ref{fig:clean_surf}. It is observed from Fig. \ref{fig6_1} that a rise in $Re$ in the surfactant-free case increases the amplitude and the frequency of the interfacial oscillations at the bubble tail; this is similar to the observations made by Magnini {\it et al.} \cite{Magnini_prf_2017} who examined the interfacial undulations of elongated `clean' bubbles in confined geometries. The accumulation of surfactants at the trailing end of the bubble (see Fig. \ref{fig6_2}) and the associated Marangoni stresses lead to dampening of these oscillations for both investigated $Re$. This effective Marangoni-induced reduction in the mean radius at the back of the bubble is accompanied by a slight increase in bubble length, which is more pronounced for the $Re=443$ case. The rigidifying effect of the Marangoni stresses also manifests itself clearly in Fig. \ref{fig6_3} through the suppression of the oscillations in the dimensionless streamwise component of the interfacial velocity, $\tilde{u}_t$, present at the back of the surfactant-free bubble; this effect is also seen in the decrease of the peak amplitude of $\tilde{u}_t$ near the bubble tip and its shift upstream. 

\subsection{Bulk surfactant effects}
\begin{figure}[htbp!] 
     \centering

     \begin{subfigure}[b]{\textwidth}
         \centering
         \includegraphics[width=1\linewidth]{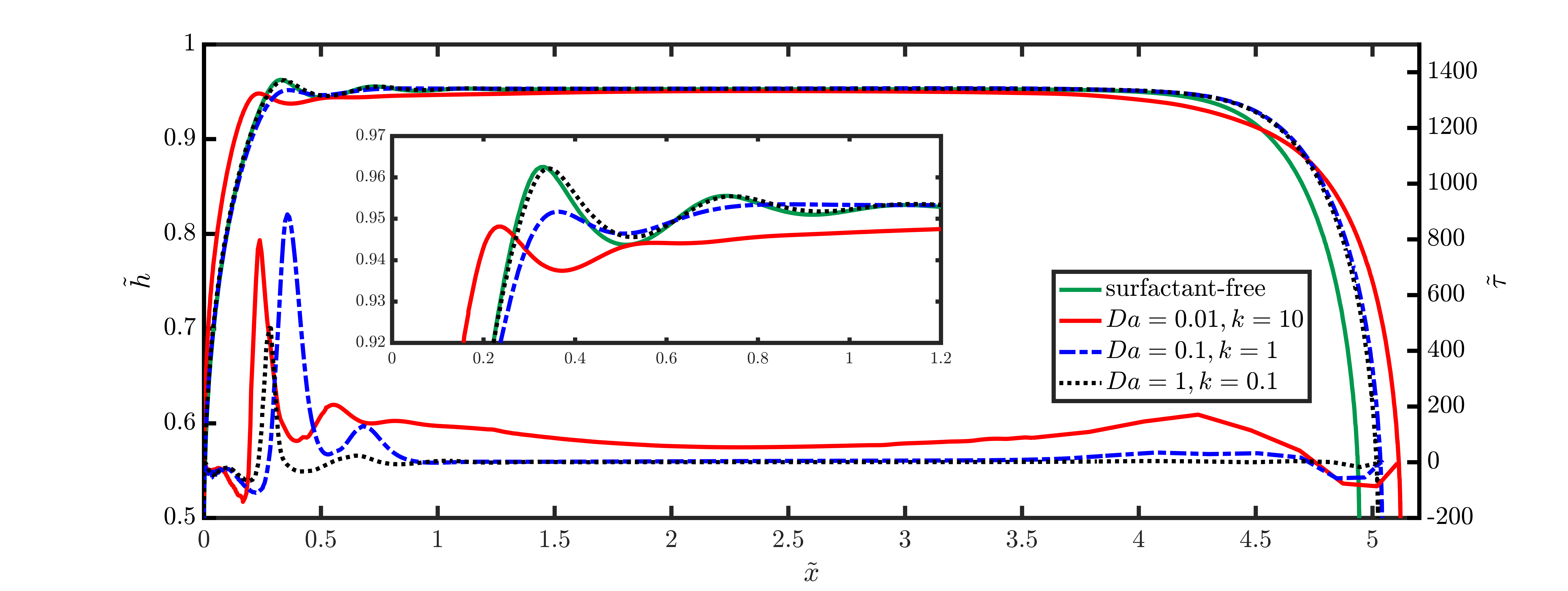}
         \caption{}
         \label{fig8_1}
     \end{subfigure}
     \hfill
     \begin{subfigure}[b]{\textwidth}
         \centering
         \includegraphics[width=1\linewidth]{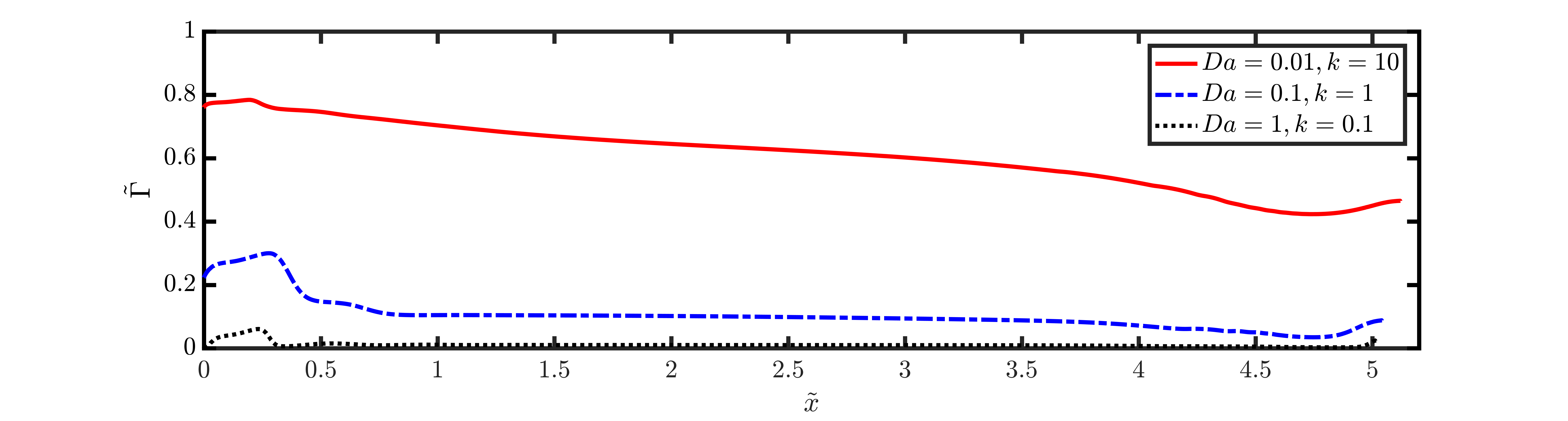}
         \caption{}
         \label{fig8_2}
     \end{subfigure}
     \hfill
    \begin{subfigure}[b]{\textwidth}
         \centering
         \includegraphics[width=1\linewidth]{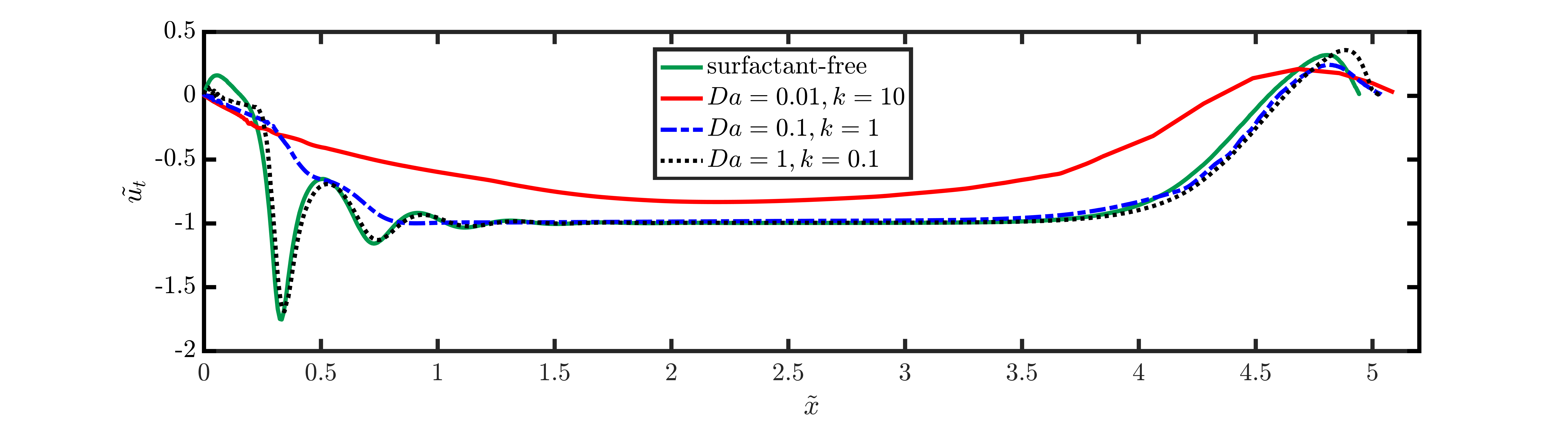}
         \caption{}
         \label{fig8_3}
     \end{subfigure}
     \hfill
        \caption{Effect of varying $Da$ and $k$ on the steady spatial distribution of the Marangoni stresses and two-dimensional projection (in the $z=0$ plane) of the bubble shape, (a), the interfacial surfactant concentration, (b), and the streamwise component of the interfacial velocity in the frame-of-reference of the bubble tip, (c), where $\tilde{u}_t=(u_t-U_b)/U_b$; here, $Ca=0.0089$, and the rest of the parameters remaining unchanged from Fig. \ref{fig:clean_surf}. 
        }
        \label{fig:Da}
\end{figure}

We now examine the effect of varying the Damkohler number, $Da$, and the surfactant adsorption depth, $k$, on the flow profiles with $Ca=0.0089$, $Re=443$, the rest of the parameters remain unchanged from Fig. \ref{fig:clean_surf}. The dimensionless group $Da$, within the context of the present work, measures the relative significance of the bulk surfactant concentration, $C_\infty$. The parameter $k$  
controls the surfactant sorption kinetics: for a fixed $C_\infty$, large values of $k$ correspond to small desorption and/or large adsorption constants, and hence slow desorption and/or rapid adsorption. In order to keep all other parameters constant, we vary $Da$ and $k$ simultaneously.
In Fig. \ref{fig:Da}, we show the bubble shape, and spatial distributions of $\tilde{\Gamma}$, $\tilde{u}_t$, and the Marangoni stresses for $Da=0.01, 0.1, 1$ with $k=10, 1, 0.1$, respectively. For $Da=1$ and $k=0.1$, 
the bubble shape and the $\tilde{u}_t$ profiles exhibit virtually no difference from the surfactant-free case since the bulk concentration is relatively low and a limited amount of surfactant remains on the interface (see Fig. \ref{fig8_2}). 
Decreasing the value of $Da$ from unity to 0.1, with $k$ increasing to 1, leads to a significant increase in $\tilde{\Gamma}$ accompanied by a rise in the magnitude of the Marangoni stresses, which result in damping of the bubble oscillations and rigidification of the tail region. A further decrease in $Da$ from 0.1 to 0.01, with $k$ increasing to 10, corresponding to an order of magnitude rise in $C_\infty$,
leads to more surfactant being adsorbed onto the interface, and a qualitative change in the structure of the $\tilde{\Gamma}$ spatial distribution. As shown in Fig. \ref{fig8_2}, although $\tilde{\Gamma}$ remains highest at the bubble tail, its distribution no longer exhibits a quasi-constant region in the middle of the bubble as had been observed in Figs. \ref{fig:clean_surf}-\ref{fig:Re}. As a result, the Marangoni stresses are non-zero over the entire bubble, even in the thin film region, which no longer has a uniform thickness but is sloped from the bubble midpoint towards the front and rear menisci (see Fig. \ref{fig8_1}); furthermore, it is seen that the bubble is elongated significantly for $Da=0.01$. Although the magnitude of the Marangoni stresses for $Da=0.01$ at the bubble tip and tail are respectively higher and lower than those associated with $Da=0.1$, the cumulative effect is a substantial reduction in the magnitude of $\tilde{u}_t$, as shown in Figs. \ref{fig8_1} and \ref{fig8_3}, i.e. the interface allows only partial slip.  

\begin{figure}[htbp!] 
     \centering
     \begin{subfigure}[b]{\textwidth}
         \centering
         \includegraphics[width=1\linewidth]{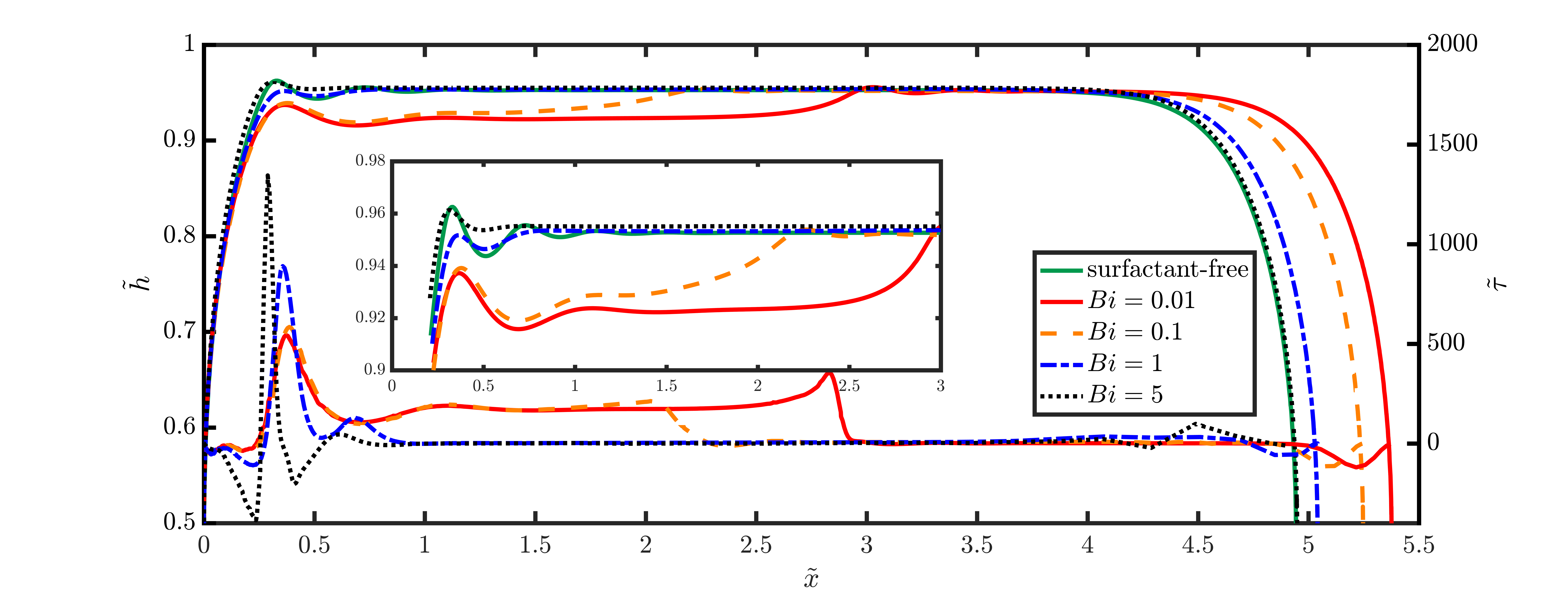}
         \caption{}
         \label{fig9_1}
     \end{subfigure}
     \hfill
     \begin{subfigure}[b]{\textwidth}
         \centering
         \includegraphics[width=1\linewidth]{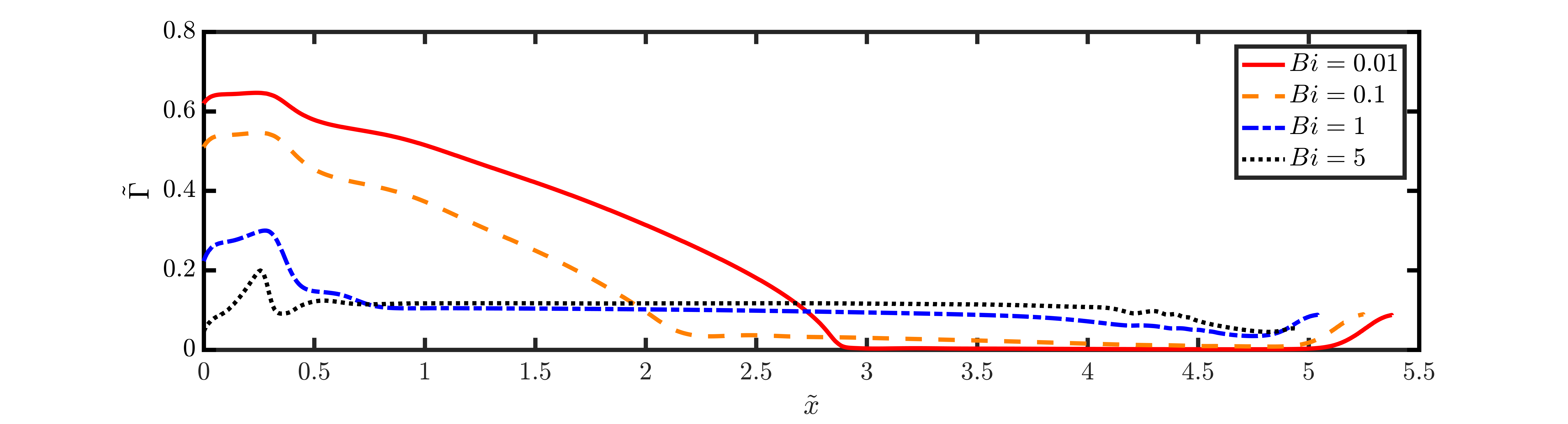}
         \caption{}
         \label{fig9_2}
     \end{subfigure}
     \hfill
    \begin{subfigure}[b]{\textwidth}
         \centering
         \includegraphics[width=1\linewidth]{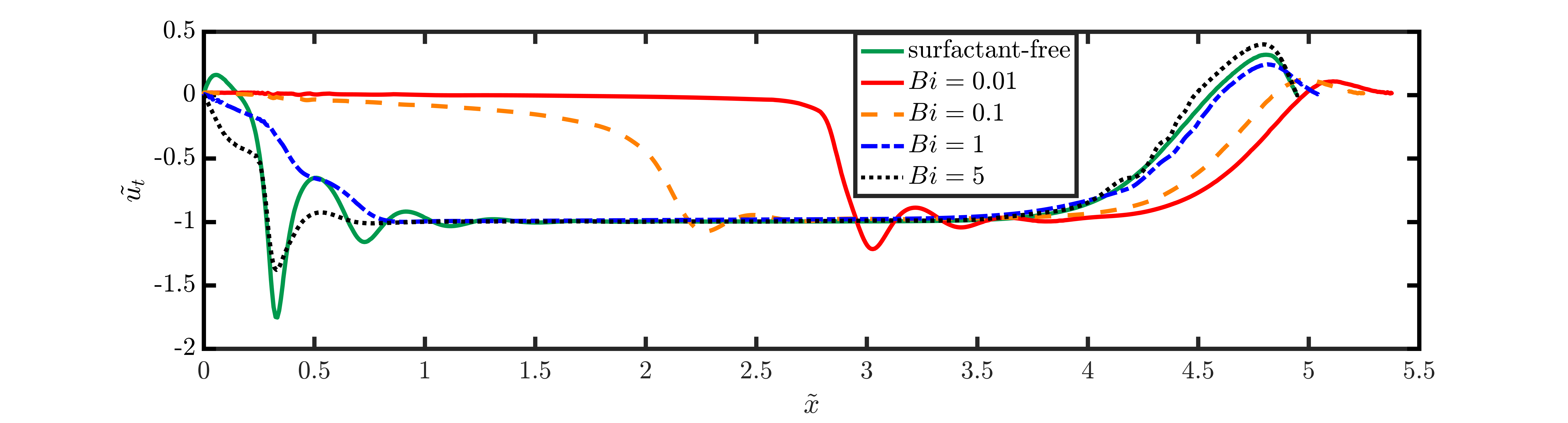}
         \caption{}
         \label{fig9_3}
     \end{subfigure}
     \hfill
        \caption{Effect of varying $Bi$ on the steady spatial distribution of the Marangoni stresses and two-dimensional projection (in the $z=0$ plane) of the bubble shape, (a), the interfacial surfactant concentration, (b), and the streamwise component of the interfacial velocity in the frame-of-reference of the bubble tip, (c), where $\tilde{u}_t=(u_t-U_b)/U_b$; here, $Ca=0.0089$, and the rest of the parameters remaining unchanged from Fig. \ref{fig:clean_surf}.
        }
        \label{fig:Bi}
\end{figure}

Next, we study the effect of the Biot number, $Bi$, which is a ratio of the flow and desorption time scales; thus, small $Bi$ values are characteristic of slow surfactant desorption kinetics. 
In Fig. \ref{fig:Bi}, which depicts the steady bubble shape and $\tilde{\Gamma}$, $\tilde{u}_t$, and Marangoni stress distributions for $Bi$ in the range $0.01-5$, it is seen clearly that this parameter has a profound effect on these profiles. In particular, there is a critical $Bi$ value that is a function of the remaining parameters, for which there is a flow regime transition. 

For $Bi=1,\,5$, all flow variables shown in Fig. \ref{fig:Bi} exhibit similar profiles to those discussed above: surfactant accumulation at the bubble rear, and Marangoni-driven rigidification leading to damping of tail oscillations. As depicted in Fig. \ref{fig9_2} for $Bi=0.1$, however, it appears that 
the bubble is divided into two distinct regions: a surfactant-covered region, Region `1', in which $\tilde{\Gamma}$ decreases from the bubble rear to very low values at the bubble midpoint approximately, which gives way to Region `2' that extends to the bubble tip with much smaller concentrations. Following the tail undulation, the liquid film in Region 1 decreases gradually to an essentially constant value, which marks the beginning of the Region 2, as shown in Fig. \ref{fig9_1}. Inspection of Fig. \ref{fig9_3} reveals that $\tilde{u}_t$ is essentially zero at the bubble rear, highlighting the rigidifying effect of the surfactant in Region 1, before reaching a value of $\tilde{u}_t=-1$, via a damped oscillation, at the start of Region 2. Due to the virtual absence of surfactant, Region 2 is considerably more mobile than Region 1 for the $Bi=1$ case. Interestingly, Region 2 also exhibits undulations at its trailing edge, which are similar to those observed at the tail of a `clean' bubble though of smaller amplitude. This is due to the sign of the interfacial curvature upstream of the undulations, which is positive in the clean bubble case, and negative at the beginning of Region 2 wherein the interface must adjust to an essentially flat Region 1.

The trends for the $Bi=0.1$ case become more pronounced by lowering $Bi$ further to $Bi=0.01$: there is a significant rise in $\tilde{\Gamma}$ in Region 1, the majority of which is rigid, and whose length is extended beyond the bubble midpoint. The transition between Regions 1 and 2 is much sharper for $Bi=0.01$ in comparison to the $Bi=0.1$ case characterised by abrupt film-thinning, rapid variation in $\tilde{u}_t$ from $\tilde{u}_t=0$ to $\tilde{u}_t=-1$, and a front-like structure exhibited by the Marangoni stress at the leading edge of Region 1. The bubble also becomes more elongated following the decrease in $Bi$.

It is worth remarking on the fact that the surfactant-laden interface becomes, effectively, a no-slip surface in Region 1 for $Bi=0.01$; this is chiefly the reason underlying the film-thickening in this region shown in Fig. \ref{fig9_1}. 
Parallels can be drawn with the work of Yu {\it et al.} \cite{Yu_2017}, where similar observations were made when the bubble rear was coated with particles. These authors found that the measured film thickness of the particle-coated thicker film region grows by a factor of $2^{2/3}$ in comparison to the solution for a `clean' bubble. \textcolor{red}{The thickening factor observed in this work in the case of $Bi=0.01$ is $1.5876$, which is approximately $2^{2/3}$.}

\begin{figure} 
     \centering

     \begin{subfigure}[b]{0.6\textwidth}
         \centering
         \includegraphics[width=0.95\linewidth]{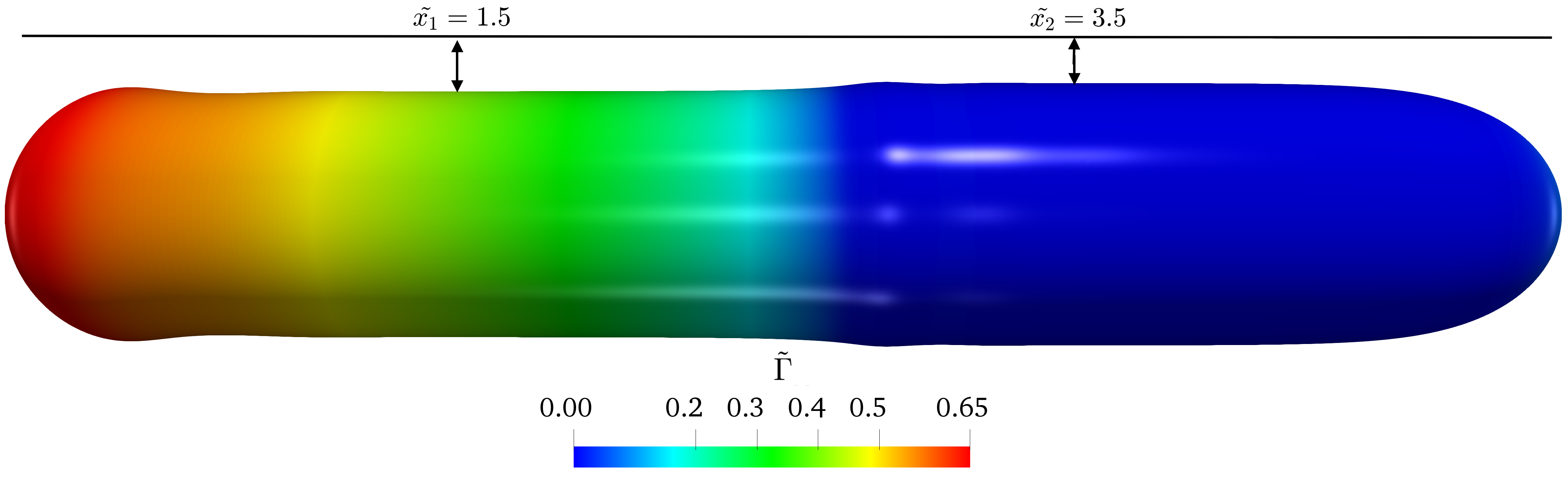}
         \caption{}
         \label{fig10_3}
     \end{subfigure}
     \hfill
     \begin{subfigure}[b]{0.3\textwidth}
         \centering
         \includegraphics[width=0.95\linewidth]{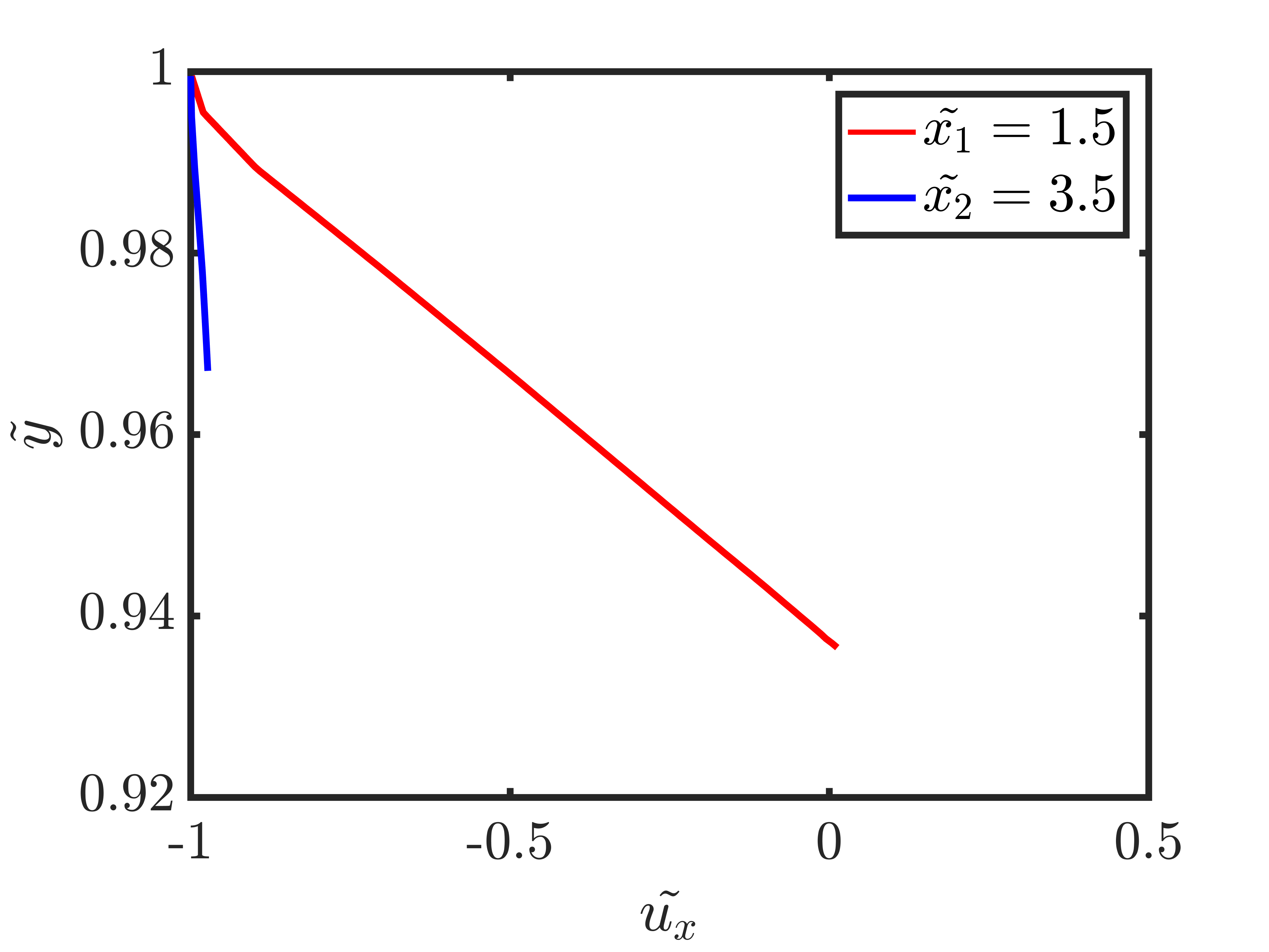}
         \caption{}
         \label{fig10_4}
     \end{subfigure}
     \hfill
     \centering
     \begin{subfigure}[b]{0.9\textwidth}
         \centering
         \includegraphics[width=1\linewidth]{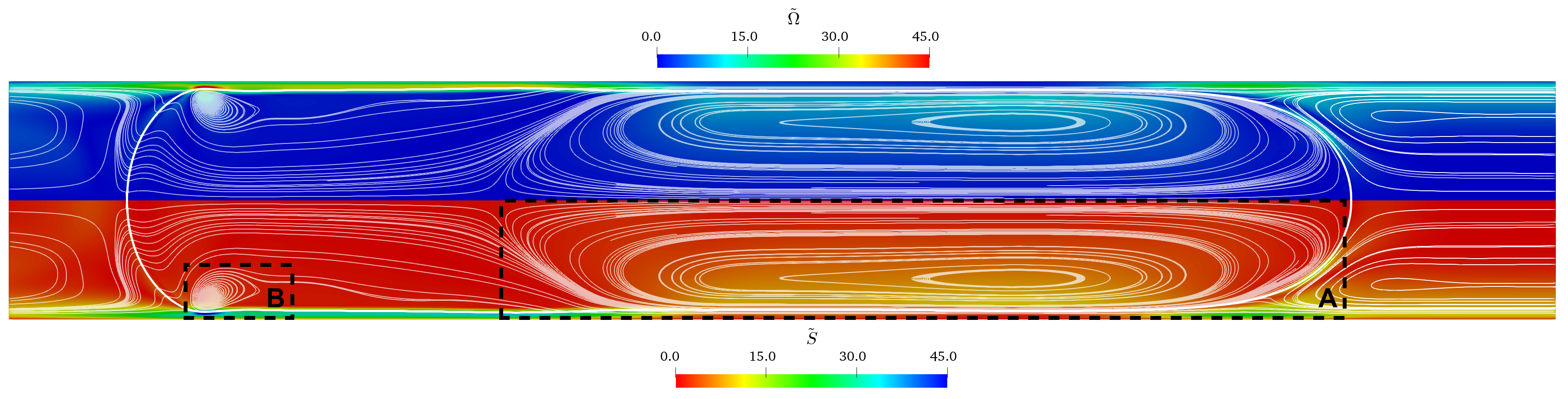}
         \caption{}
         \label{fig11_1}
     \end{subfigure}
     \hfill
     \begin{subfigure}[b]{0.9\textwidth}
         \centering
         \includegraphics[width=1.0\linewidth]{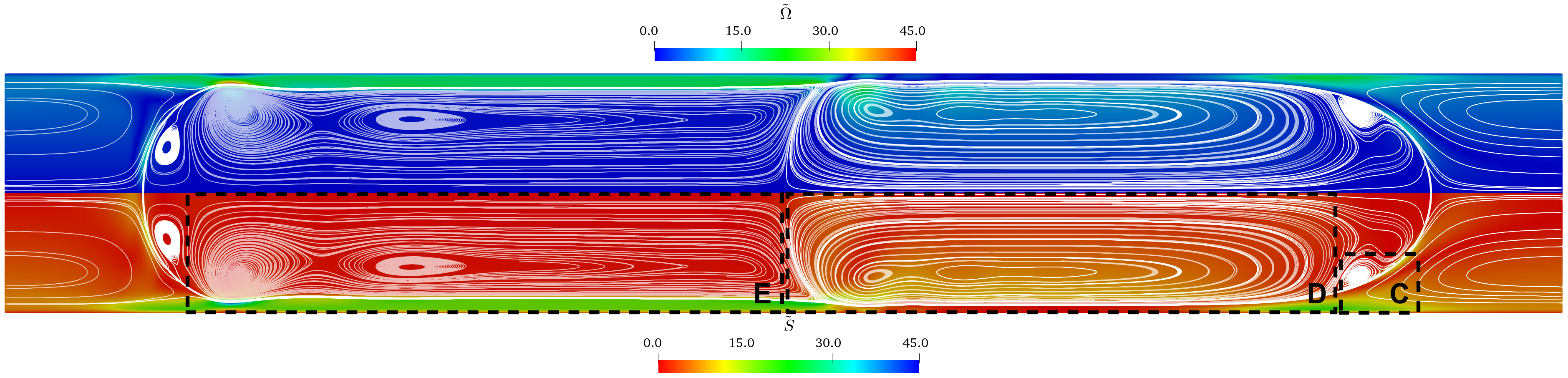}
         \caption{}
         \label{fig11_2}
     \end{subfigure}
     \hfill
        \caption{Flow profiles associated with the $Bi=0.01$ case with the rest of the parameters remaining unchanged from Fig. \ref{fig:Bi}; (a) three-dimensional representation of the bubble shape coloured by the magnitude of the interfacial surfactant distribution, $\tilde{\Gamma}$; (b) profiles of the dimensionless streamwise velocity component, $\tilde{u}_x=(u_x-U_b)/U_b$, calculated in a bubble-tip frame-of-reference, along the cross-stream direction, $y$, for the two axial locations indicated in (a), which are in Regions 1 and 2, as described in the text; vortical structure evolution at $\tilde{t}=3$ and at steady-state, shown in (c) and (d), with the magnitude of the dimensionless vorticity, $\tilde{\Omega}$, and strain rate, $\tilde{S}$ depicted in the top-half and bottom-half of each panel, respectively. The vortices labelled `A' and `B' in panel (c) rotate in the clockwise and   counter-clockwise directions, respectively. In panel (d), the vortices labelled `C' and `E', and vortex `D' rotate in the counter-clockwise, and clockwise directions, respectively; identical structures in the top half of the panel rotate in the opposite directions. All streamlines are presented in a frame-of-reference moving with the bubble tip. }
        \label{fig:film velocity}
\end{figure}

In Fig. \ref{fig10_3} we show a three-dimensional representation of the bubble shape for the $Bi=0.01$ case with the colour being indicative of the magnitude of $\tilde{\Gamma}$; this shows clearly the surfactant-laden and surfactant-free regions discussed above. We also plot in Fig. \ref{fig10_4} the variation of the dimensionless streamwise velocity component, $\tilde{u}_x$, in the wall-normal direction within the films in Regions 1 and 2 in a frame-of-reference moving with the bubble. It is seen that in Region 1, $\tilde{u}_x=-1$ and zero at the tube wall and the gas-liquid interface, respectively, which correspond to no-slip conditions reflecting the rigidified nature of the interface in this region. As a result, the $\tilde{u}_x$ profile in Region 1 is predominantly Couette-like due to the absence of significant pressure gradients arising from interfacial curvature effects. In the more mobile Region 2, the liquid in this region is effectively in plug flow since $\tilde{u}_x=-1$ at the wall located at $\tilde{y}=1$, and the interfacial condition corresponds, effectively, to one of no-shear stress, $\partial \tilde{u}_x/\partial \tilde{y} \approx 0$, due to the absence of surfactant.

The Couette-like profile in Region 1 leads to the development of a zone with nearly-uniform vorticity and strain rates across this region, as shown in Fig. \ref{fig11_1} and \ref{fig11_2}. Exploring the evolution of the vortical structures, it is seen in Fig. \ref{fig11_1} that prior to the development of Region 1 two counter-rotating re-circulation zones form at the front and the back of the bubble (see zones `A' and `B' in Fig. \ref{fig11_1}). The vortex identified in zone `A' helps the migration of surfactant species towards the bubble tail, whereas the one at the bubble tail (see zone `B' in Fig. \ref{fig11_1}) inhibits the surfactants from migrating further back. This gives rise to elevated Marangoni stresses, which as seen leads to the creation of the thicker film region by pushing the liquid towards the centre of the tube. At steady-state, the larger vortex that spans across the thicker film region (see zone `E' in Fig. \ref{fig11_2}) is counter-rotating to the one ahead of it (see zone `D' in Fig. \ref{fig11_2}), helping the preservation of a constant vorticity rate across that region. An additional vortex forms at the bubble head promoting the migration of surfactant species towards the bubble tip (see zone `C' in Fig. \ref{fig11_2}). 


\begin{figure} [htbp!] 
     \centering

     \begin{subfigure}[b]{\textwidth}
         \centering
         \includegraphics[width=1\linewidth]{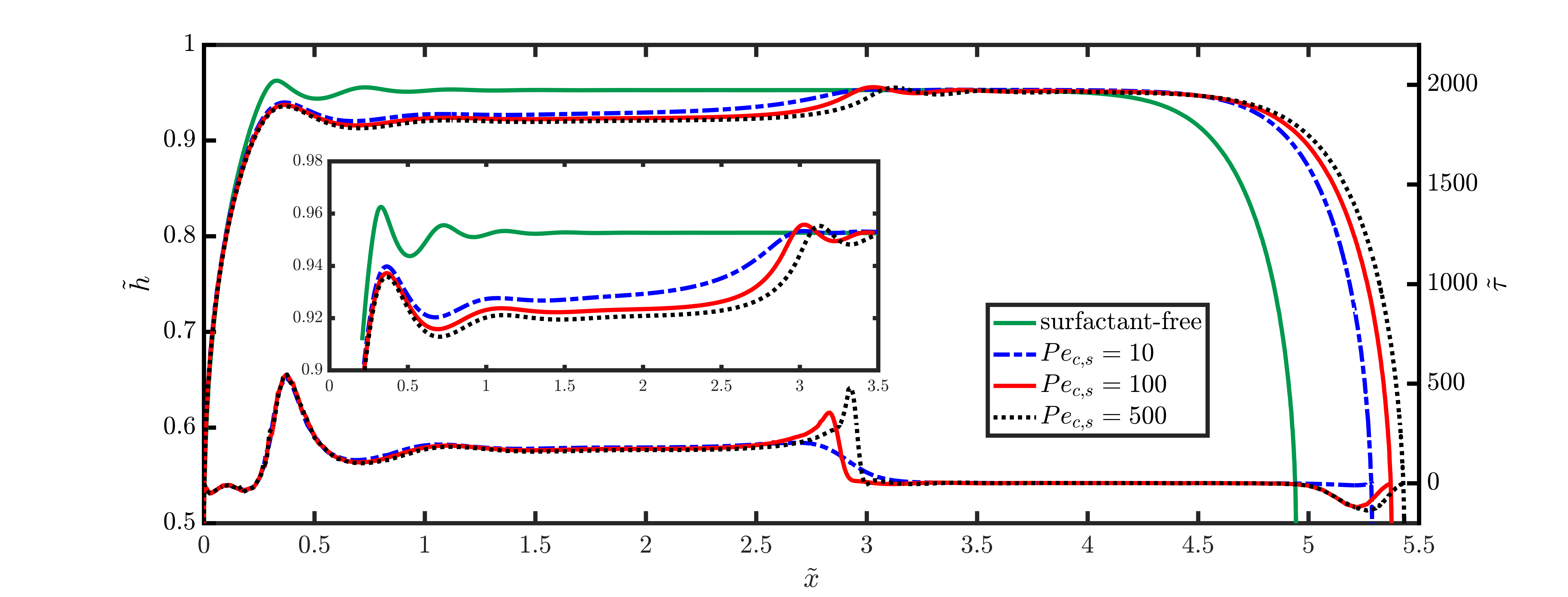}
         \caption{}
         \label{fig12_1}
     \end{subfigure}
     \hfill
     \begin{subfigure}[b]{\textwidth}
         \centering
         \includegraphics[width=1\linewidth]{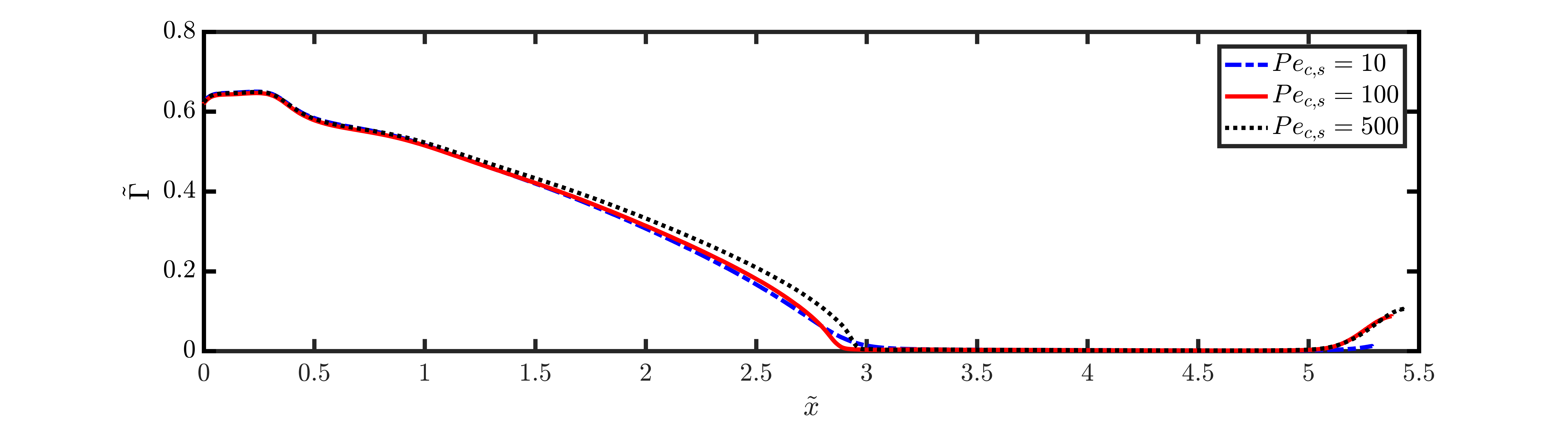}
         \caption{}
         \label{fig12_2}
     \end{subfigure}
     \hfill
    \begin{subfigure}[b]{\textwidth}
         \centering
         \includegraphics[width=1\linewidth]{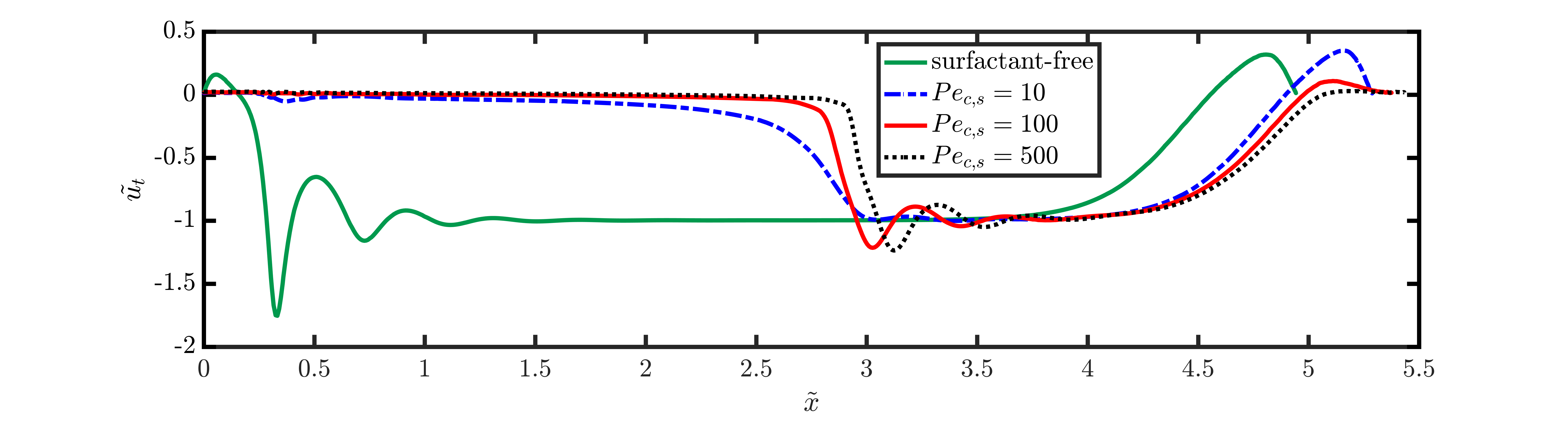}
         \caption{}
         \label{fig12_3}
     \end{subfigure}
     \hfill
        \caption{Effect of varying $Pe_c$ (set equal to $Pe_s$) on the steady spatial distribution of the Marangoni stresses and two-dimensional projection (in the $z=0$ plane) of the bubble shape, (a), the interfacial surfactant concentration, (b), and the streamwise component of the interfacial velocity in the frame-of-reference of the bubble tip, (c), where $\tilde{u}_t=(u_t-U_b)/U_b$; here, $Ca=0.0089$, $Bi=0.01$, and the rest of the parameters remain unchanged from Fig. \ref{fig:clean_surf}.
        }
        \label{fig:PeBi}
\end{figure}

We now examine the effect of diffusion, whose relative significance is characterised by the interfacial and bulk Peclet numbers, on the steady flow profiles of the $Bi=0.01$ case; the results are shown in Fig. \ref{fig:PeBi} with $Pe_c=Pe_s$, and the rest of the parameters remaining unchanged from Fig. \ref{fig:Bi}. First we inspect the effect of $Pe_{c,s}$ on the surfactant-covered Region 1. Upon investigation of the bubble shape in Fig. \ref{fig12_1}, we see that Region 1 is the thickest and longest one observed for the highest investigated Peclet number (i.e. $Pe_{c,s}=500$). Lowering bulk and interfacial diffusion allows for the Marangoni stress field to push further towards the bubble tip in comparison to $Pe_{c,s}=100$. In addition, inspection of Fig. \ref{fig12_2} reveals a steeper concentration gradient between Regions 1 and 2 for $Pe_{c,s}=500$ in comparison to $Pe_{c,s}=100$, whereas, for $Pe_{c,s}=10$, the spatial distribution of $\tilde{\Gamma}$ is more gradual between the two regions, highlighting the stronger diffusive effects.  
The elimination of the abrupt concentration gradient in the case of $Pe_{c,s}=10$, results in the suppression of undulation structures at the beginning of Region 2. In Fig. \ref{fig12_3}, we observe that the mobility of this zone is also reduced.

\subsection{Effect of bubble length at low $Bi$}
\begin{figure}[t!]  
    \centering
    \begin{subfigure}[b]{1\textwidth}
    \includegraphics[width=1\linewidth]{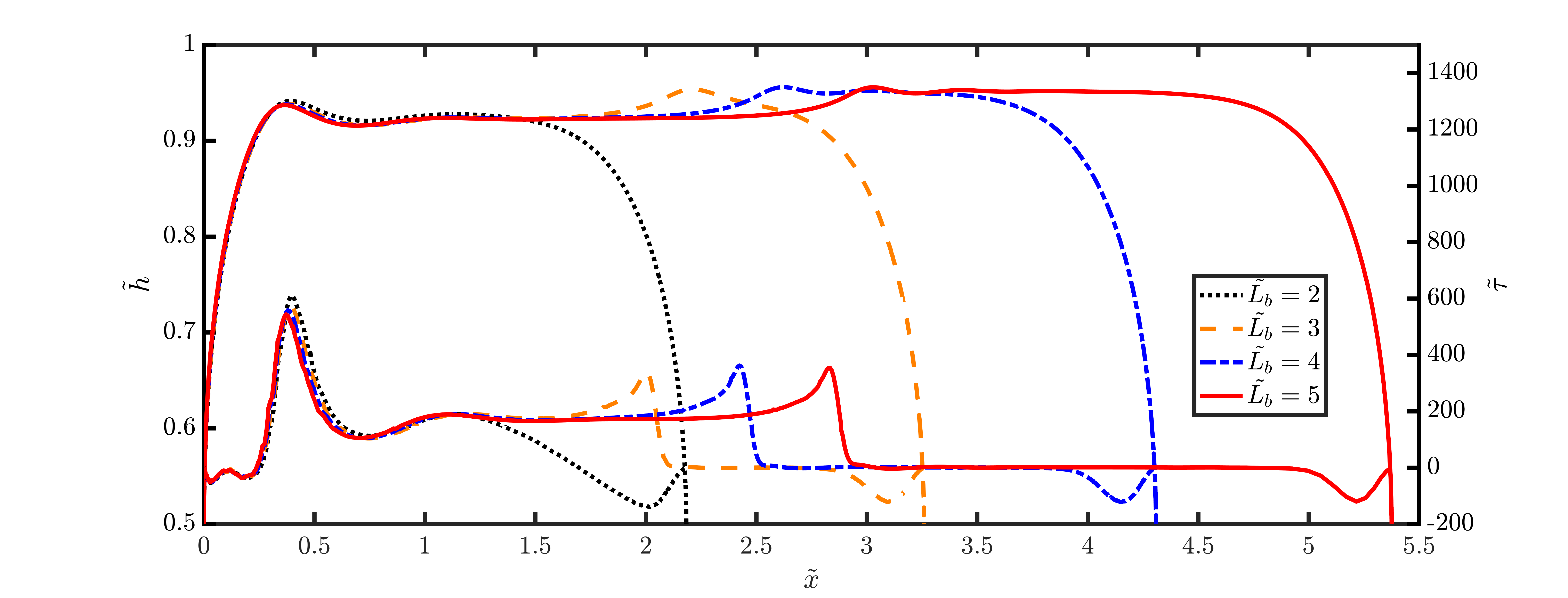}
    \caption{}
    \label{fig16_1}
    \end{subfigure}
    \hfill
        \begin{subfigure}[b]{1\textwidth}
    \includegraphics[width=1\linewidth]{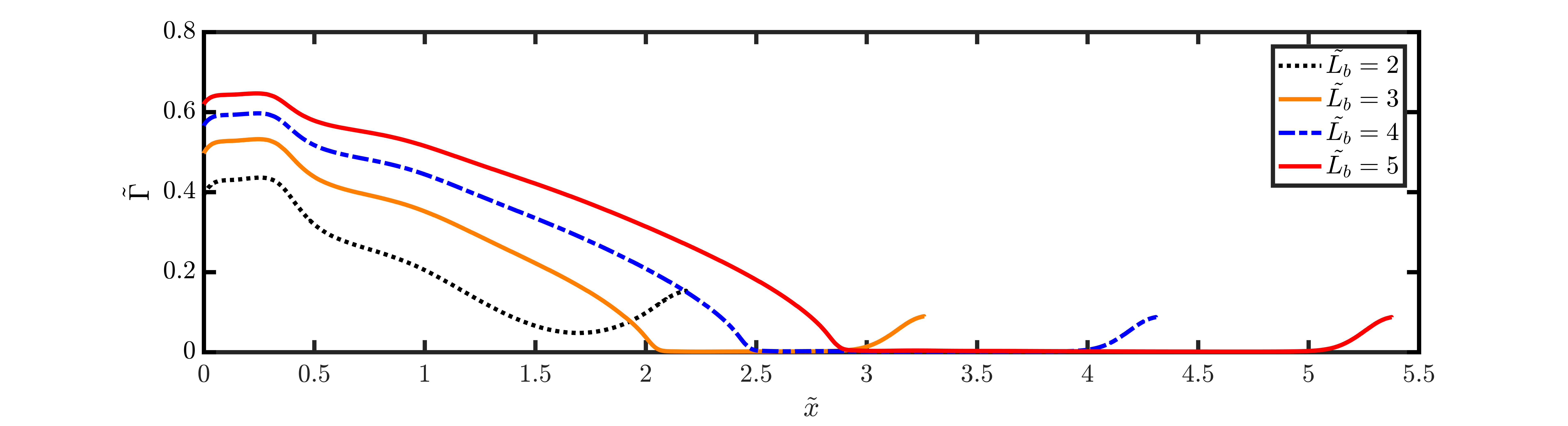}
    \caption{}
    \label{fig16_2}
    \end{subfigure}
    \hfill
        \begin{subfigure}[b]{1\textwidth}
    \includegraphics[width=1\linewidth]{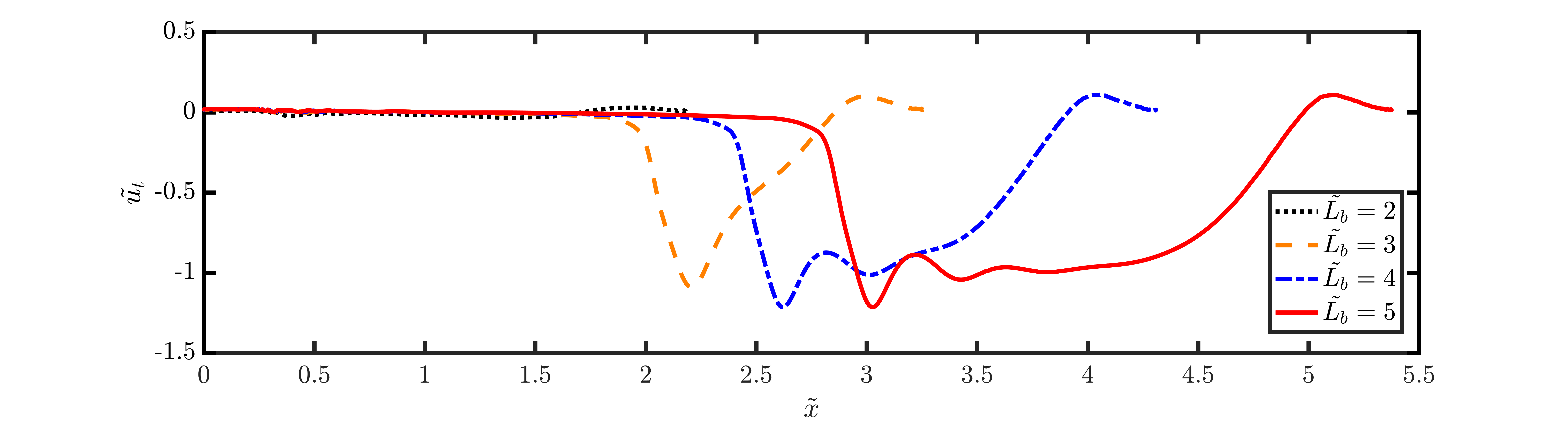}
    \caption{}
    \label{fig16_3}
    \end{subfigure}
    \hfill
    \caption{Effect of varying the initial dimensionless bubble length, $\tilde{L}_b$ on the steady 
    spatial distribution of the Marangoni stresses and two-dimensional projection (in the $z=0$ plane) of the bubble shape for $\tilde{L}_b=2-5$, (a), the interfacial surfactant concentration, (b), and the streamwise component of the interfacial velocity in the frame-of-reference of the bubble tip, (c), where $\tilde{u}_t=(u_t-U_b)/U_b$; here, $Bi=0.01$, $Ca=0.0089$, and the rest of the parameters remaining unchanged from Fig. \ref{fig:clean_surf}.
    }
    \label{fig:length} 
\end{figure}

Finally, we study the effect of varying the dimensionless bubble length, $\tilde{L_b}\equiv L_b/D$, on the flow profiles for the $Bi=0.01$ case
with $Ca=0.0089$ and the rest of the parameters remaining unaltered from Fig. \ref{fig:clean_surf}. We focus on the influence of $\tilde{L_b}$ on the development of the two regions that arise for sufficiently low $Bi$ values discussed above.  
A summary of the results is shown in Fig. \ref{fig:length}. 
It is seen clearly from Fig. \ref{fig16_1} that for the shortest bubbles examined, the interface is covered fully with surfactant, with the peak of the distribution located near the bubble rear. The Marangoni stresses associated with this case act to smooth the tail oscillations and `rigidify' the interface effectively, as indicated by the low value of $\tilde{u}_t$ presented in Fig. \ref{fig16_3}. Furthermore, the two-region structure observed for low Biot numbers discussed in the previous section is absent in the $\tilde{L}_b=2$ case. In contrast, the remaining cases examined, for which $\tilde{L}_b=3-5$, all demonstrate the development of a trailing edge, the shape \textcolor{red}{and length} of which is governed by the elevated Marangoni stress zone, and ahead of which there is a markedly thinner surfactant-free region, as shown in Fig. \ref{fig16_1}. Though the latter region becomes more fully-developed with increasing $\tilde{L}_b$, there is very little qualitative difference between the flow profiles associated with these $\tilde{L}_b=3-5$ cases. 
\section{Conclusion \label{sec:conc}}
The effect of surfactants on the dynamics of elongated bubbles propagating through capillary tubes was studied extensively using a hybrid front-tracking/level-set method. The convective-diffusive transport of surfactant species along the gas-liquid interface and in the bulk is fully-coupled to the Navier-Stokes equations, where surface tension is related to the interfacial surfactant concentration using a non-linear Langmuir  equation of state. The simulations performed in this work consider the effects of inertia, capillarity, bulk and interfacial diffusion, and Marangoni stresses arising from the presence of surfactant-induced surface tension gradients, on the flow dynamics; attention was focused on high Reynolds numbers. The numerical predictions were validated against previous experimental work \cite{Han_ijhff_2009} before performing a full parametric study. 

It was found that the presence of surfactants is  effective in suppressing the bubble tail undulations, which are otherwise present in surfactant-free systems. In addition, at the lower range of the  capillary and Reynolds numbers examined, surfactants are found to have a wall-film thickening effect, attributed to the formation of Marangoni stresses.
We have also shown that increasing the strength of Marangoni stresses 
reduces the mobility of the interface at the bubble rear significantly, and at high bulk concentrations these stresses influence the flow profiles over the entire bubble. 
At low surfactant solubility, characterised by small Biot numbers, the steady bubble shape features the formation of two distinct regions: a surfactant-covered, interfacially-immobile region at the bubble rear, and another, mobile region downstream extending to the bubble tip. The thick liquid film associated with the former region gives way to a thinner film via a transition region whose length decreases with decreasing Biot and increasing Peclet numbers. Connections are established with previous studies involving armored confined bubbles, where the bubble rear is covered with colloidal particles \cite{Yu_2017}, in which similar phenomena are observed. 

Finally, the effect of bubble length in the context of low solubility surfactants was also investigated. It was shown that almost complete rigidification of the bubble interface occurs for bubbles with sufficiently small initial length, beyond which 
qualitatively similar profiles were observed, characterised by the formation of the two-region structures mentioned above.
\subsection*{Acknowledgements}
This work is supported by the Engineering $\&$ Physical Sciences Research Council, United Kingdom, through the MEMPHIS (EP/K003976/1) and PREMIERE (EP/T000414/1) Programme Grants, and by computing time at HPC facilities provided by the Research Computing Service (RCS) of Imperial College London.  The numerical simulations were performed with code BLUE (Shin et al. \cite{Shin_jmst_2017}) and the visualisations have been generated using ParaView. The authors wish to thank with gratitude Dr T. Abadie and G. F. N. Gon\c{c}alves for meaningful discussions.


\begin{thebibliography}{1}


\bibitem{Aussillous_pof_2000}
P. Aussillous and  D. Qu\'er\'e,
\newblock Quick deposition of a fluid on the wall of a tube,
\newblock {\em Phys. Fluids} \textbf{12}(10), 2367-2371 (2000).


\bibitem{Borhan_PofA_1992}
 A. Borhan and C.-F. Mao,
\newblock Effect of surfactants on the motion of drops through circular tubes,
\newblock {\em Phys. Fluids A} \textbf{4}, 2628 (1992).
 

\bibitem{Borhani_ijmf_2014}
N. Borhani and J. R. Thome,
Intermittent dewetting and dryout of annular flows,
\newblock {\em Int. J. Multiph. Flow} \textbf{12}(10), 2367-2371 (2014).


\bibitem{Bretherton_jfm_1961}
F. P. Bretherton,
\newblock The motion of long bubbles in tubes,
\newblock {\em J. Fluid Mech.} \textbf{10}(2), 166-188 (1961).


\bibitem{Campana_pof_2010}
D. M. Campana, S. Ubal, M. D. Giavedoni, and F. A. Saita,
Numerical prediction of the film thickening due to surfactants in the Landau-Levich problem,
\newblock {\em Phys. Fluids} \textbf{22}(3), 032103 (2010).

\bibitem{deRyck_pof_2002}
A. de Ryck,
\newblock The effect of weak inertia on the emptying of a tube,
\newblock {\em Phys. Fluids} \textbf{14}(7), 2102-2108 (2002).


\bibitem{Edvinsson_aiche_1996}
R. K. Edvinsson and S. Irandoust,
\newblock Finite-element analysis of Taylor flow,
\newblock {\em AIChE J.}, \textbf{42}(7), 1815-1823 (1996).


\bibitem{Ghadiali_jfm_2003}
S. N. Ghadiali and D. P. Gaver,
The influence of non-equilibrium surfactant dynamics on the flow of a semi-infinite bubble in a rigid cylindrical capillary tube,
\newblock {\em J. Fluid Mech.} \textbf{478}, 165-196 (2003).


\bibitem{Giavedoni_pof_1997}
M. D. Giavedoni and F. A. Saita,
\newblock The axisymmetric and plane cases of a gas phase steadily displacing a Newtonian liquid-A simultaneous solution of the governing equations,
\newblock {\em Phys. Fluids} \textbf{9}(8), 2420-2428 (1997).


\bibitem{Giavedoni_pof_1999}
M. D. Giavedoni and F. A. Saita,
\newblock The rear meniscus of a long bubble steadily displacing a Newtonian liquid in a capillary tube,
\newblock {\em Phys. Fluids} \textbf{11}(4), 786-794 (1999).


\bibitem{Ginley_acsss_1989}
G. M. Ginley and C. J. Radke,
Influence of Soluble Surfactants on the Flow of Long Bubbles Through a Cylindrical Capillary,
\newblock {\em ACS Symp. Ser.} \textbf{396}(26), 480-501 (1989).


\bibitem{Grotberg_arbe_2001}
J. B. Grotberg,
\newblock Respiratory Fluid Mechanics and Transport Processes,
\newblock {\em Annu. Rev. Biomed. Eng.} \textbf{3}, 421-457 (2001).


\bibitem{Grotberg_pof_2011}
J. B. Grotberg.
\newblock Respiratory Fluid Mechanics.
\newblock {\em Phys. Fluids} \textbf{23}, 021301 (2011).


\bibitem{Han_ijhff_2009}
Y. Han and N. Shikazono,
\newblock Measurement of the liquid film thickness in micro tube slug flow,
\newblock {\em Int. J. Heat Fluid Flow} \textbf{30}(5), 842-853 (2009).


\bibitem{Halpern_rpn_2008}
D. Halpern, H. Fujioka, S. Takayama, and J. B. Grotberg,
Liquid and surfactant delivery into pulmonary airways,
\newblock {\em Respir. Physiol. Neurobiol.} \textbf{163}(1-3), 222-231 (2008).


\bibitem{Heil_rpn_2008}
M. Heil,  A. L. Hazel, and  J. A. Smith,
The mechanics of airway closure,
\newblock {\em Respir. Physiol. Neurobiol.} \textbf{163}(1-3), 214-221 (2008).


\bibitem{Heil_pof_2001}
M. Heil,
Finite Reynolds number effects in the Bretherton problem,
\newblock {\em Phys. Fluids} \textbf{13}(9), 2517-2521 (2001).

\bibitem{Johnson_jciS}
R. A. Johnson and A. Borhan,
Pressure-driven motion of surfactant-laden drops through cylindrical capillaries: effect of surfactant solubility,
\newblock {\em J. Colloid Interface Sci.} \textbf{261}(2), 529–541 (2003).

\bibitem{Kahouadji_mano_2018}
L. Kahouadji, E. Nowak, J. Kovalchuk, S. Shin, J. Chergui, D. Juric, M. Simmons, R. V. Craster, and O. K. Matar,
\newblock  Simulation of immiscible liquid-liquid flows in complex microchannel geometries using a front-tracking scheme,
\newblock {\em Microfluid. Nanofluid.}, \textbf{22}(11), 126 (2018).



\bibitem{Khodaparast_mano_2015}
S. Khodaparast, M. Magnini, N. Borhani, and J. R. Thome,
Dynamics of isolated confined air bubbles in liquid flows through circular microchannels: an experimental and numerical study,
\newblock {\em Microfluid. Nanofluid.} \textbf{19}(1), 209-234 (2015).


\bibitem{Khodaparast_est_2017}
S. Khodaparast, M. Kevin Kim, J. E. Silpe, and H. A. Stone,
Bubble-Driven Detachment of Bacteria from Confined Microgeometries,
\newblock {\em 	Environ. Sci. Technol.} \textbf{51}(3), 1340--1347 (2017).


\bibitem{Krechetnikov_pof_2005}
R. Krechetnikov and G. M. Homsy,
Experimental study of substrate roughness and surfactant effects on the Landau-Levich law,
\newblock {\em Phys. Fluids} \textbf{17}, 102108 (2005).


\bibitem{Kreutzer_aiche_2005}
M. T. Kreutzer, F. Kapteijn, J. A. Moulijn,  C. R. Kleijn, and  J. J. Heiszwolf,
Inertial and interfacial effects on pressure drop of Taylor flow in capillaries,
\newblock {\em AIChE J.}, \textbf{51}(9), 2428-2440 (2005).


\bibitem{Magnini_ijts_2016}
M. Magnini and J. R. Thome,
A CFD study of the parameters influencing heat transfer in microchannel slug flow boiling,
\newblock {\em Int. J. Therm. Sci.} \textbf{110}, 119-136 (2016).


\bibitem{Magnini_prf_2017}
M. Magnini, A. Ferrari,  J. R. Thome, and H. A. Stone.
Undulations on the surface of elongated bubbles in confined gas-liquid flow.
\newblock {\em Phys. Rev. Fluids} \textbf{2}(8), 1-21 (2017)


\bibitem{Matar_sf_2009}
O. K. Matar and R. V. Craster.
Dynamics of surfactant-assisted spreading.
\newblock {\em Soft Matter} \textbf{5}(20), 3801-3809 (2009)

\bibitem{Muradoglu_jcp_2014}
M. Muradoglu and G. Tryggvason,
Simulations of soluble surfactants in 3D multiphase flow,
\newblock {\em J. Comp. Phys.} \textbf{274}, 737-757 (2014).


\bibitem{Olgac_ijmf_2013}
U. Olgac and M. Muradoglu,
Effects of surfactant on liquid film thickness in the Bretherton problem,
\newblock {\em Int. J. Multiph. Flow} \textbf{48}, 58-70 (2013).


\bibitem{OuRamdane_Langmuir_1997}
O. Ou Ramdane and D. Qu\'er\'e,
Thickening Factor in Marangoni Coating,
\newblock {\em Langmuir} \textbf{13}(11), 2911-2916 (1997).


\bibitem{Park_pofA_1992}
C-W. Park,
Influence of soluble surfactants on the motion of a finite bubble in a capillary tube,
\newblock {\em Phys. Fluids A} \textbf{4}, 2335 (1992).


\bibitem{Ratulowski_jfm_1990}
J. Ratulowski and H.-C. Chang,
Marangoni effects of trace impurities on the motion of long gas bubbles in capillaries,
\newblock {\em J. Fluid Mech.} \textbf{210}, 303-328 (1990).


\bibitem{Russell_ces_2019}
A. W. Russell, L. Kahouadji, K. Mirpuri, A. Quarmby, P. M. Piccione, O. K. Matar, P. F. Luckham, and C. N. Markides,
Mixing viscoplastic fluids in stirred vessels over multiple scales: A combined experimental and CFD approach,
\newblock {\em Chem. Eng. Sci.} \textbf{208} 115129 (2019).


\bibitem{Severino_pof_2003}
M. Severino, M. D. Giavedoni, and F. A. Saita,
A gas phase displacing a liquid with soluble surfactants out of a small conduit: The plane case,
\newblock {\em Phys. Fluids} \textbf{15}, 2961 (2003).


\bibitem{Shin_jcp_2002}
S. Shin and D. Juric,
\newblock  Modelling three-dimensional multiphase flow using a level contour reconstruction method for front tracking without connectivity,
\newblock {\em J. Comp. Phys.}, \textbf{180} (2), 427 - 470 (2002).


\bibitem{Shin_jcp_2005}
S. Shin, D. Juric, S, Abdel-Khalik, and V. Daru,
\newblock  Accurate representation of surface tension using the level contour reconstruction method,
\newblock {\em J. Comp. Phys.}, \textbf{203} (2), 493 - 516 (2005).


\bibitem{Shin_jmst_2007}
S. Shin and D. Juric,
\newblock  High order level contour reconstruction method,
\newblock {\em J. Mech. Sci. Technol.}, \textbf{21} (2), 311-326 (2007).


\bibitem{Shin_jmst_2017} 
S. Shin, J. Chergui, and D. Juric,  
\newblock  A solver for massively parallel direct numerical simulation of three-dimensional multiphase flows,
\newblock {\em J. Mech. Sci. Technol.}, \textbf{31}, 1739-175 (2017).


\bibitem{Shin_jcp_2018}
S. Shin, J. Chergui, D. Juric, L. Kahouadji, O. K. Matar, and R. V.  Craster,
\newblock   A hybrid interface tracking – level set technique for multiphase flow with soluble surfactant,
\newblock {\em J. Comp. Phys.}, \textbf{359}, 409 - 435 (2018).


\bibitem{Stebe_jfm_1995}
K. J. Stebe and D. Barth\`es-Biesel,
Marangoni effects of adsorption-desorption controlled surfactants on the leading end of an infinitely long bubble in a capillary,
\newblock {\em J. Fluid Mech.} \textbf{286}, 25-48 (1995).


\bibitem{Taylor_jfm_1960}
G. I. Taylor,
\newblock Deposition of a viscous fluid on a plane surface, 
\newblock {\em J. Fluid Mech.} \textbf{9}(2), 218-224 (1960).


\bibitem{Yu_2017}
Y. Yu, S. Khodaparast, and H. A. Stone,
\newblock Armoring confined bubbles in the flow of colloidal suspensions, 
\newblock {\em Soft Matter} \textbf{13}(15), 2857-2865 (2017).


 \end{thebibliography}
 \end{document}